\documentclass[prl,twocolumn,showpacs]{revtex4}
\usepackage{graphicx,amsfonts,amssymb,amsmath}

\begin{document}

\title{Capillary-Wave Model for the Solidification of Dilute Binary Alloys}

\date{\today}

\author{Alexander L. Korzhenevskii}\affiliation{Institute for Problems of Mechanical Engineering, RAS, Bol'shoi prospect. V. O., 61, St Petersburg, 199178, Russia}
\author{Richard Bausch}\affiliation{Institut f{\"u}r Theoretische Physik IV,Heinrich-Heine-Universit{\"a}t D{\"u}sseldorf, Universit{\"a}tsstrasse 1, D-40225 D{\"u}sseldorf, Germany}
\author{Rudi Schmitz}\affiliation{Institut f{\"u}r Theoretische Physik A, RWTH Aachen, Templergraben 55, D-52056 Aachen, Germany}

\date{\today}

\begin{abstract}

Starting from a phase-field description of the isothermal solidification of a dilute binary alloy, we establish a model where capillary waves of the solidification front interact with the diffusive concentration field of the solute. The model does not rely on the sharp-interface assumption, and includes non-equilibrium effects, relevant in the rapid-growth regime. In many applications it can be evaluated analytically, culminating in the appearance of an instability which, interfering with the Mullins-Sekerka instability, is similar to that, found by Cahn in grain-boundary motion. 

\end{abstract}

\maketitle

\section{Introduction}

Crystal growth from an under-cooled fluid phase is frequently described by a diffusion equation for heat or particle density, complemented by boundary conditions at a moving sharp interface between the solid and fluid phases. As explained in the review \cite{Langer-RMP} by Langer, one of the boundary conditions comprises the conservation of the density field in terms of the related flux and source terms at the interface. A second boundary condition is the Gibbs-Thomson relation which is adequate in cases of local thermodynamic equilibrium. For applications in the rapid-growth regime Aziz and Boettinger \cite{AB} have extended the Gibbs-Thomson relation by including non-equilibrium effects which they derived in an intermediate step from an atomistic picture of the initially extended interface region.

One of the most exciting achievements of the sharp-interface approach is the explanation of dendritic growth in the diffusion-limited regime, opening up at low under-cooling. The initiation of this process is described by the Mullins-Sekerka instability \cite{MS} whereas the cumbersome route to an analytic calculation of the fully-developed three-dimensional dendrite by Brener \cite{Brener-PRL} is reviewed in Ref. \cite{Brener}. Another exciting case of pattern formation in the kinetics-limited rapid-growth regime is the periodic formation of layers with alternating homogeneous and dendritic micro-structures in dilute binary alloys \cite{CGZK}. A crucial point in explanations of this effect by Carrard et al. \cite{CGZK}, and by Karma and Sarkissian \cite{KS} is the non-monotonous dependence of the interface temperature on the growth velocity, observed in Ref. \cite{AB}.    

A fundamentally different access to a theory of crystal growth rests on the use of a phase-field order parameter which allows a continuous transition between the solid and fluid phases. This approach is closely related to that, used by Halperin, Hohenberg, and Ma \cite{HHM} for studying dynamic critical behavior in their model C , and, in the context of crystal-growth processes, has been described by Collins and Levine \cite{Collins}, by Caginalp and Fife \cite{Caginalp}, and by Langer \cite{Langer-Dir}. Phase-field models for binary alloys have been established by Wheeler et al. \cite{Wheeler}, and later by Kim et al. \cite{Kim}. In the article \cite{Langer-Dir} by Langer it is emphasized that one reason for promoting the phase-field approach is the hope, that similar to the identification of universality classes for dynamic critical phenomena, one may obtain a fundamental understanding of various pattern-forming mechanisms in solidification problems. A quite different reason for the interest in the phase-field approach is that it presents a convenient basis for simulations, avoiding the costly interface-tracking procedure, necessary in the sharp-interface description.

Within a phase-field approach L\"owen et al. \cite{LBT} have investigated the long-time crossover behavior between the diffusion- and the kinetics-limited regime. One of their results is the appearance of meta-stable branches where the steady-state motion of a planar solidification front can occur in the diffusion-limited regime. These branches are part of the trajectories which show the non-monotonous dependence of the interface temperature on velocity, later found in Ref. \cite{AB}. In a phase-field model of rapid solidification Ahmad et al. \cite{Ahmad} have considered the effects of solute trapping and solute drag, and found partial agreements with the results, obtained in Ref. \cite{AB}. A good example for the computational use of the phase-field approach is the work of Karma and Rapell \cite{KR} on a quantitative description of dendritic growth. In their treatment the above-mentioned non-monotonic velocity dependence of the interface temperature is noticed by the appearance of a negative kinetic coefficient, joining the term, linear in velocity. Since such an anomaly does not arise in the sharp-interface description of dendritic growth, the authors invented a compensation device for the unwanted term.

Among the numerous papers on derivations of sharp-interface descriptions from a phase-field model the work by Elder et al. \cite{Elder} is probably the most elaborate one. Their approach avoids the assumption of a zero interface width, but instead uses the products of this width with the interface curvature, and with the growth rate divided by the diffusion constant, as small expansion parameters. Since, however, the second parameter is of order one in the rapid-growth regime, applications to this regime are excluded. The result for the Gibbs-Thomson relation in the low-velocity regime contains a kinetic-under-cooling term where, as in Ref. \cite{KR}, the kinetic coefficient can, in some parameter range, become negative, an effect which regrettably has not been scrutinized by the authors.

In the present paper we advertise a model description where, retaining a finite interface width, the interface position is used as basic field variable, in addition to a bulk-diffusion field. Although the energy density is a feasible example of a bulk field, we will predominantly consider the isothermal solidification of a dilute binary alloy, in which this role is taken by the concentration of the solute component. An advantage of our approach is that it remains valid in the rapid-growth regime which can be seen in a thorough derivation from a phase-field model. Our description can, however, also be derived in a self-contained way from first principles, in the first line from the observation that the presence of an interface breaks Euclidean symmetry, and, therefore, implies the existence of Goldstone modes \cite{GNW}. These are capillary waves of the solidification front which interact with the bulk-diffusion mode. The interaction kernel as well as the diffusion coefficient can be freely chosen in our approach which, therefore, can be applied to a sizable variety of model systems.

For a subset of models with a location-independent diffusion constant, we have derived a universal form of the dispersion relation of interface eigenmodes. These modes determine all possible morphological instabilities of the solidification front, and in all cases we encounter the Mullins-Sekerka instability. It is the only instability which survives the sharp-interface limit in our scheme. As soon as we allow a finite interface width, we become aware of an additional instability which features a finite amplification rate already for a planar perturbation, a behavior, previously discovered by Cahn \cite{Cahn} in grain-boundary motion. The nature of this instability is closely related to the non-monotonic velocity dependence of the interface temperature. In our approach the shift of this temperature from the melting point of the solvent enters as a driving force which we also evaluated analytically for several models. Although these subjects are, like the distantly-related Corriel-Sekerka instability \cite{CS}, mostly discussed in connection with the rapid-growth regime, they also affect the low-velocity behavior, as illustrated by the previously mentioned sign problem of the kinetic coefficient.

One of the toy models, investigated in the frame of our capillary-wave approach, allows a surprisingly simple analysis of some attributes of the solidification process. In this model both input-functions, the solute-interface interaction and the solute-diffusion coefficient, are taken to linearly interpolate between the solid and fluid bulk phases, assuming a finite interface width. The effect of solute trapping is measured by the partition coefficient, for which we found an expression, coinciding with that, derived by Aziz and Kaplan \cite{AK}, up to some re-scaling of the involved characteristic velocity. We, furthermore, have established a simple analytic relation between the driving force and the growth rate which reflects the non-monotonous behavior of the interface temperature. The minima of all driving-force trajectories are connected by a kinetic spinodal line which we also have determined analytically, finding agreement with an analogous line in the non-isothermal growth of a one-component crystal, discussed by Umantsev \cite{Umantsev}. The image of this spinodal line in the temperature-concentration phase diagram is located between the solidus line and a static spinodal line, established by Baker and Cahn \cite{Baker}. Within our model this provides an answer to an issue concerning this matter, discussed by Hillert in Ref. \cite{Hill}.

In our concluding discussion we present an expression for the entropy production in the steady-state growth of a planar solidification front. The result demonstrates that the appearance of a negative kinetic coefficient, also in our approach, is not in conflict with basic principles of linear irreversible thermodynamics. Also included in our discussion are some estimates for the relevant model parameters which determine the range of validity of our approach. We, finally, emphasize the flexibility of our description, concerning generalizations and the inclusion of additional field variables.

\section{Phase-Field Description}

In order to derive our capillary-wave model from a standard phase-field model, we initially describe the isothermal solidification of a dilute binary alloy by a phase field $\Phi({\bf r},t)$ for the solvent, and a concentration field $\mathcal{C}({\bf r},t)$ of the solute component. In terms of these field variables the effective Hamiltonian of our model reads

\begin{center}
\begin{figure}
 \includegraphics[width=8cm]{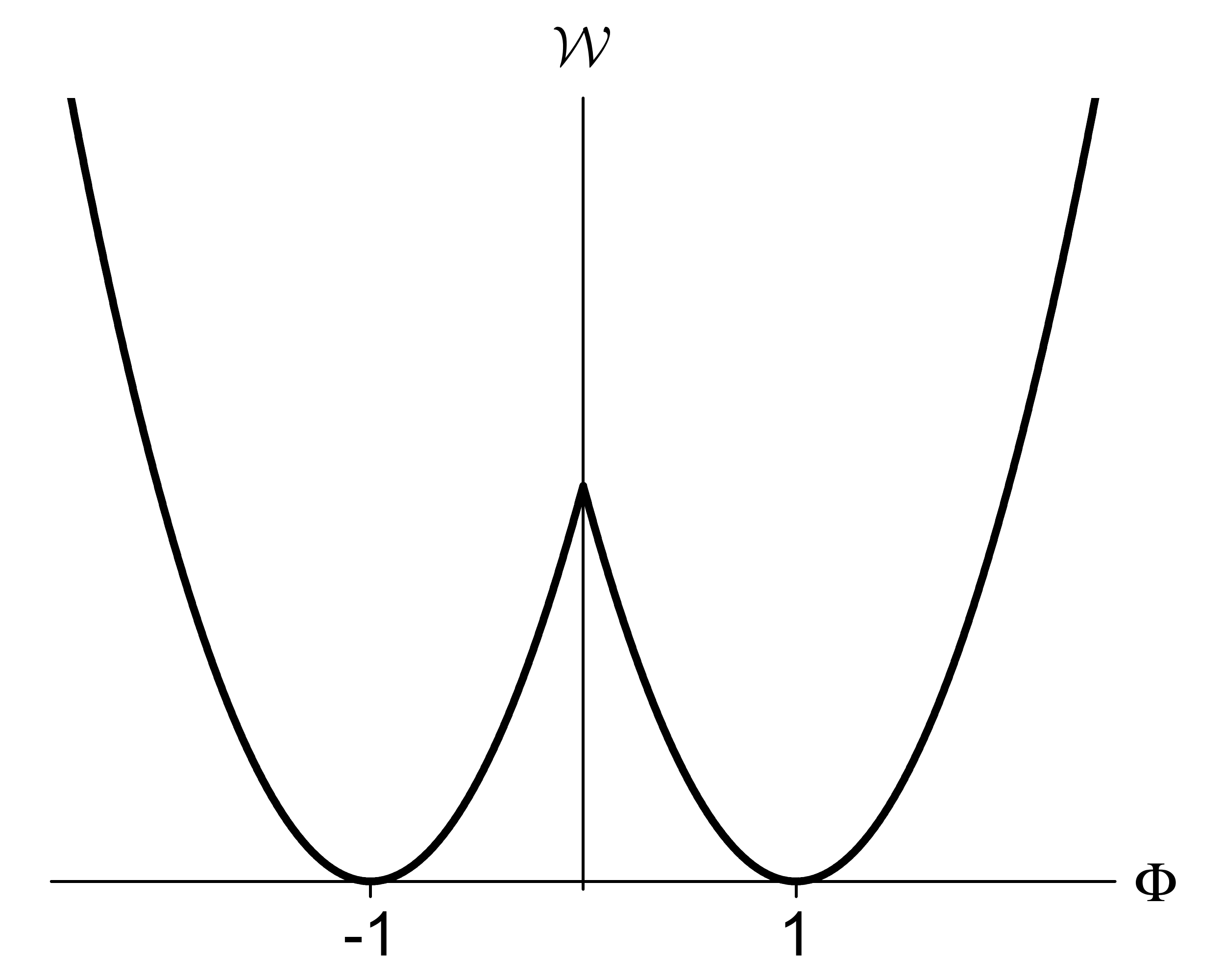}
 \caption{\label{double-well}{Double-parabola potential $\mathcal{W}(\Phi)$ with equilibrium phase-field values $\Phi=+1$ in the liquid, and $\Phi=-1$ in the solid phase.}}
 \end{figure}
 \end{center} 

\begin{eqnarray}\label{Hamiltonian}
\mathcal{H}=\int d^3r\biggl\{\frac{\sigma}{\xi}\left[\frac{\xi^2}{2}(\nabla\Phi)^2+\mathcal{W}(\Phi)\right]\biggr.&&
\nonumber\\+\biggl.\frac{\kappa(T)}{2}\,[\,\mathcal{C}-\mathcal{U}(\Phi,T)]^2\biggr\}&&.
\end{eqnarray}
Here, $\mathcal{W}(\Phi)$ is a double-well potential with degenerate minima at $\Phi=+1$ and $\Phi=-1$, representing the liquid and solid equilibrium phases of the solvent. In the present paper we will exclusively use the explicit form 

\begin{equation}\label{doublewell}
\mathcal{W}(\Phi)=\Theta(-\Phi)\,\frac{1}{2}\,(\Phi+1)^2+\,\Theta(\Phi)\,\frac{1}{2}\,(\Phi-1)^2\,\,,
\end{equation}
shown in Fig. 1. Contrary to that, the potential $\mathcal{U}(\Phi,T)$ is adaptable, up to the relations

\begin{equation}\label{C-L,C-S}
\mathcal{U}(-1,T)=C_S(T)\,\,\,,\,\,\,\mathcal{U}(+1,T)=C_L(T)
\end{equation}
where $C_S(T)$ and $C_L(T)$ are the solute concentrations in the solid and liquid phases at temperature $T$, depicted in the temperature-concentration phase diagram in Fig. 2.

\begin{center}
\begin{figure}
 \includegraphics[width=8cm]{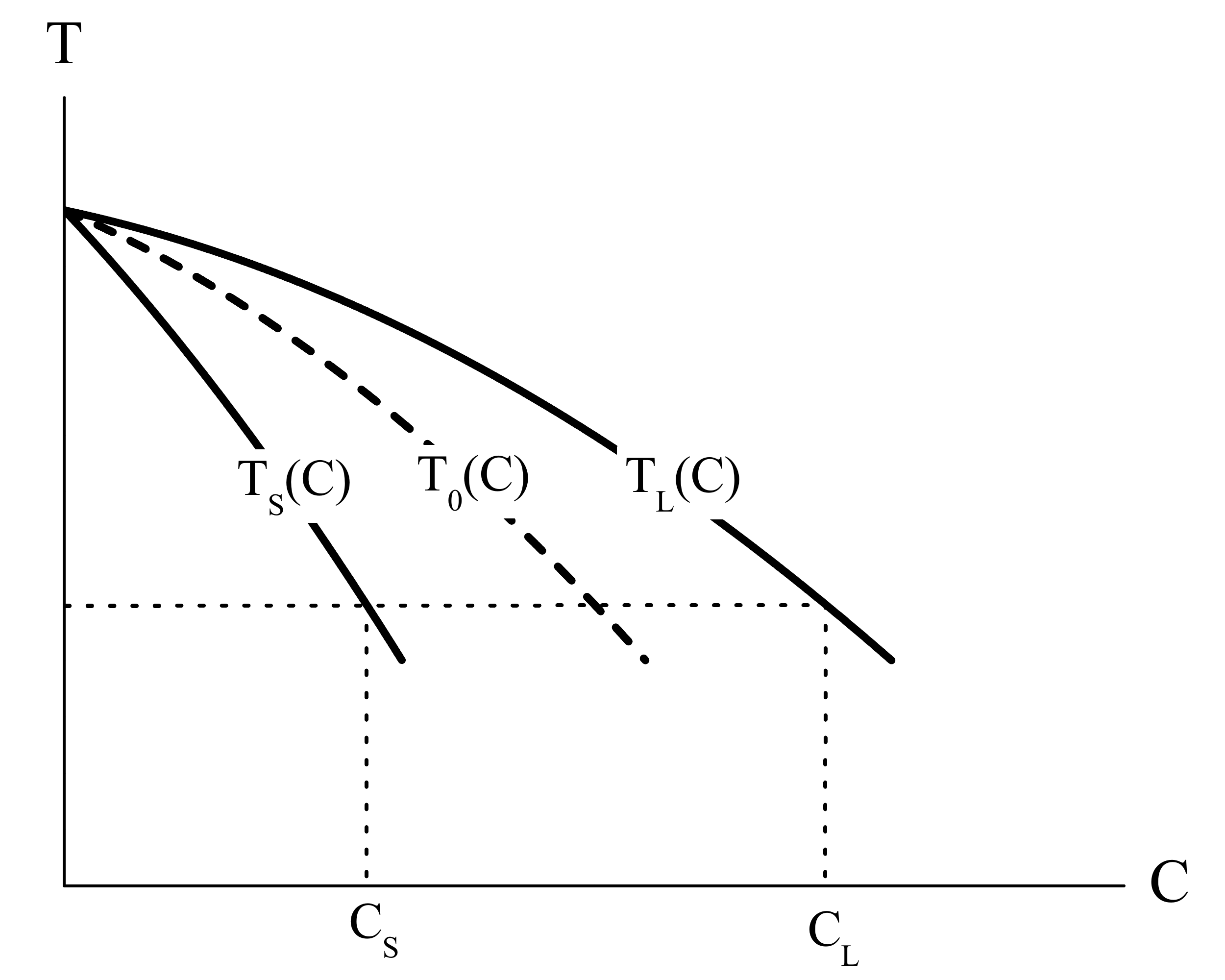}
 \caption{\label{T-C-diagram}{Temperature-concentration phase diagram, showing the liquidus and solidus lines $T_L(C),T_S(C)$ which enclose the two-phase region. The values $C_L$ and $C_S$ define the miscibility gap at the temperature $T_L(C_L)=T_S(C_S)$. Also shown is the equilibrium spinodal line $T_0(C)$ according to Baker and Cahn.}}
 \end{figure}
 \end{center}
 
\begin{center}
\begin{figure}
 \includegraphics[width=8cm]{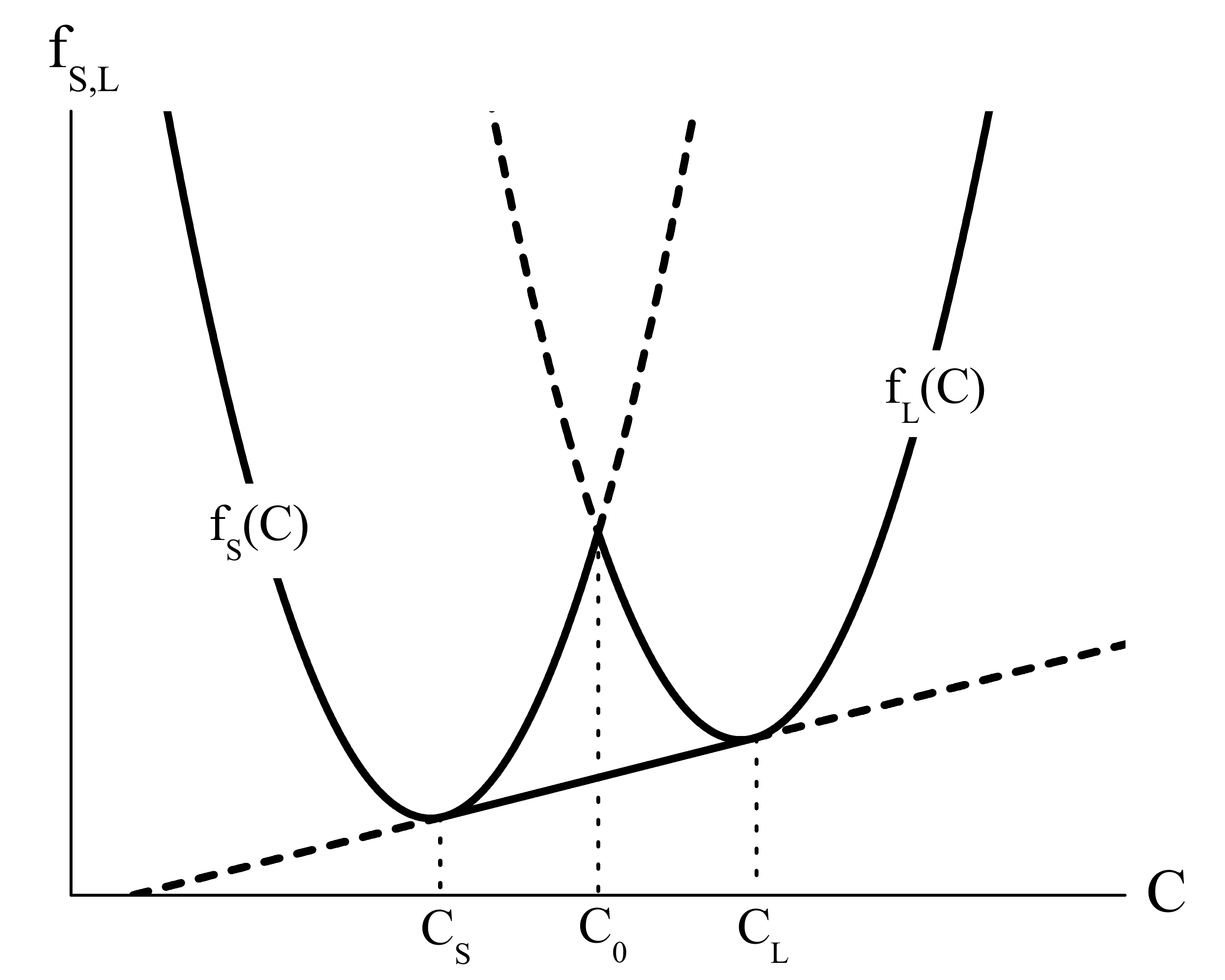}
 \caption{\label{double-tangent}{Double-tangent construction, applied to the free-energy densities $f_S(C,T)$ and $f_L(C,T)$. The intersection point $C_0(T)$ defines the $T_0$-line in Fig. 2.}}
 \end{figure}
 \end{center} 

The values (\ref{C-L,C-S}) follow from the mean-field free-energy densities

\begin{eqnarray}\label{free-energies}
f_S(C,T)&=&\frac{\kappa(T)}{2}[C-U(-1,T)]^2+\mu_E(T)C\,\,,\nonumber\\
f_L(C,T)&=&\frac{\kappa(T)}{2}[C-U(+1,T)]^2+\mu_E(T)C\,\,,
\end{eqnarray}
by the double-tangent construction, visualized in Fig. 3. In this procedure the terms, involving the equilibrium chemical potential $\mu_E(T)$, cancel. The intersection point in Fig. 3 at  

\begin{equation}\label{intersection}
C_0(T)=\frac{C_S(T)+C_L(T)}{2}
\end{equation} 
borders the regions where the two bulk phases can exist in a meta-stable state. In Fig. 2 the inverse function $T_0(C)$ defines a kind of spinodal line, promoted by Baker and Cahn \cite{Baker}. Within the often-used approximation of a constant miscibility gap this line has the form

\begin{equation}\label{T_0-line}
T_0(C)=\frac{T_S(C)+T_L(C)}{2}\,\,.
\end{equation}

The equilibrium chemical potential $\mu_E(T)$ in Eqs. (\ref{free-energies}) obeys the Clausius-Clapeyron equation 

\begin{equation}\label{Claus-Clap}
T_M\,\frac{d\mu_E}{dT}=-\frac{L}{\Delta C(T)}
\end{equation}
where $T_M$ is the melting temperature of the pure solvent, $L$ the latent heat, and

\begin{equation}\label{misc-gap}
\Delta C(T)\equiv C_L(T)-C_S(T)\,\,,
\end{equation}
is the miscibility gap. Following Langer \cite{Langer-RMP}, we can use Eq. (\ref{Claus-Clap}) to extract an expression for $\kappa(T)$ by forming the total derivative of the chemical potential 

\begin{equation}\label{chem-pot}
\mu_L(C,T)=\kappa(T)[C-C_L(T)]+\mu_E(T)
\end{equation}
with respect to temperature, taken at $C=C_L(T)$. The result reads  

\begin{equation}\label{kappa(T)}
\kappa(T)=-\left(\frac{\partial C_L}{\partial T}\right)^{-1}\frac{L}{T_M}\,\frac{1}{\Delta C(T)}
\end{equation}
where, as already in Eq. (\ref{Claus-Clap}), only leading terms in $\Delta C$ have been taken into account. We later will realize that the assumption of a small miscibility gap is a crucial assumption in the derivation of an interface description from a phase-field model.

The equilibrium conditions 

\begin{equation}\label{equilibrium}
\frac{\delta\mathcal{H}}{\delta\Phi}=0\,\,\,,\,\,\,\frac{\delta\mathcal{H}}{\delta\mathcal{C}}=0
\end{equation}
have the single-kink solution 

\begin{eqnarray}\label{kink}
\Phi_E(z)&=&\Theta(-z)[-1+\exp{(z/\xi)}]\nonumber\\&&+\Theta(z)[1-\exp{(-z/\xi)}]\,\,, 
\end{eqnarray}
describing the solid-liquid phase-field profile, displayed in Fig. 4, and an attached solute-concentration profile

\begin{equation}\label{concentration}
C_E(z)=\mathcal{U}[\Phi_E(z)]
\end{equation}
which leads to identify $2\xi$ with the width of the interface, and the parameter $\sigma$ in the Hamiltonian (\ref{Hamiltonian}) with the surface tension.

For the dynamics of the system we adopt the equations of motion 

\begin{center}
\begin{figure}
 \includegraphics[width=8cm]{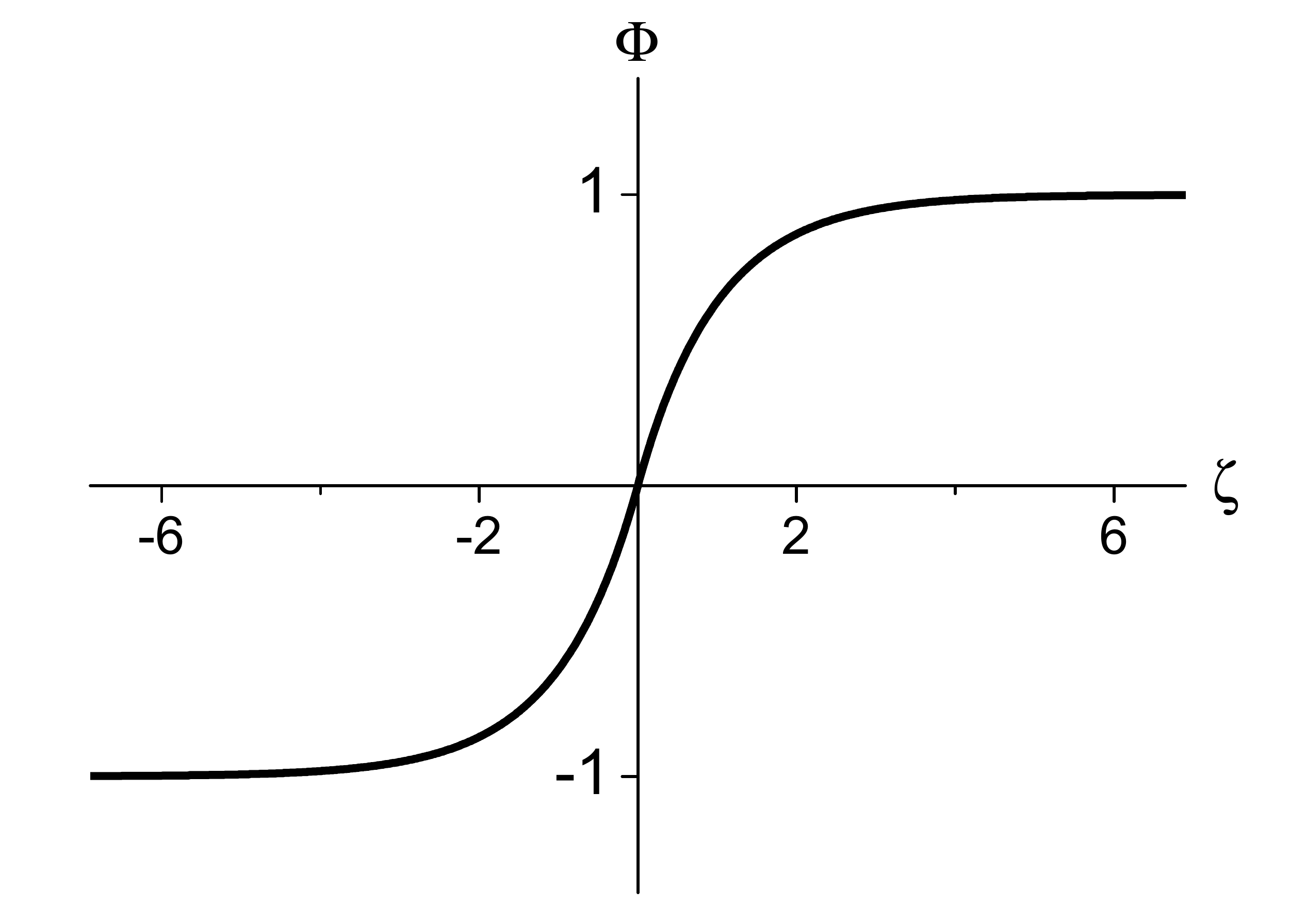}
 \caption{\label{single-kink}{Single-kink phase-field profile in thermal equilibrium, following from the potential $\mathcal{W}(\Phi)$.}}
 \end{figure}
 \end{center}

\begin{eqnarray}\label{dynamics}
\partial_t\Phi&=&-\Gamma\,\frac{\xi}{\sigma}\left[\frac{\delta\mathcal{H}}{\delta\Phi}+\mathcal{F}\right]
\,\,,\nonumber\\\partial_t\mathcal{C}&=&\nabla\cdot\mathcal{D}(\Phi)\nabla\frac{1}{\kappa}\, \frac{\delta\mathcal{H}}{\delta\mathcal{C}}
\end{eqnarray}
where $\Gamma$ is the phase-field relaxation rate, and $\mathcal{D}(\Phi)$ the diffusion coefficient of the solute atoms. The relations

\begin{equation}\label{D-L,D-S}
\mathcal{D}(+1)=D_L\,\,\,,\,\,\,\mathcal{D}(-1)=D_S
\end{equation}
allow in general different values $D_L$ and $D_S$ in the liquid and solid phases. Furthermore,

\begin{equation}\label{drive}
\mathcal{F}(\Phi)=F[\Theta(-\Phi)(\Phi+1)-\Theta(\Phi)(\Phi-1)]
\end{equation}
is a driving force which is operating near the interface, and, for $F>0$, supports the solid at the expense of the fluid phase, thus activating the solidification process. The form (\ref{drive}) can be viewed as arising from a potential

\begin{eqnarray}\label{double-well}
\mathcal{W}_F(\Phi)&=&\Theta(-\Phi)\,\frac{1}{2}\,\left[\left(1+\frac{\xi}{\sigma}F\right)(\Phi+1)^2
-\frac{\xi}{\sigma}F\right]\\&&+\,\Theta(\Phi)\,\frac{1}{2} \,\left[\left(1-\frac{\xi}{\sigma}F\right)(\Phi-1)^2+\frac{\xi}{\sigma}F\right]\,\,,\nonumber
\end{eqnarray}
replacing $\mathcal{W}(\Phi)$ in the Hamiltonian (\ref{Hamiltonian}). A convenience of this modelling is that it remains meaningful even close to the apparent spinodal point at $F=\sigma/\xi$, noted in Ref. \cite{SO}. We finally mention that Langevin forces, representing thermal-noise effects, have been ignored in Eqs. (\ref{dynamics}), because we are primarily interested in the non-equilibrium effects, induced by the force $\mathcal{F}$.

A dimensionless form of our model equations can be established by the mappings

\begin{eqnarray}\label{rescaling}
&&\frac{1}{\xi}\,{\bf r}\rightarrow{\bf r}\,\,\,,\,\,\,\frac{D_L}{\xi^2}\,t\rightarrow t\,\,\,,\,\,\, \frac{1}{D_L}\,\mathcal{D}\rightarrow\mathcal{D}\,\,,\\&&\frac{2}{\Delta C}(\mathcal{C}-C_S)\rightarrow \mathcal{C}\,\,\,,\,\,\,\frac{2}{\Delta C}\,\mathcal{U}\rightarrow\mathcal{U}\,\,\,,\,\,\, \frac{\xi}{\sigma}\,F\rightarrow F\nonumber
\end{eqnarray}
where the shift of the field $\mathcal{C}$ has been made with regard to the case of an under-cooling at constant concentration $C_S$. The equations of motion (\ref{dynamics}) then assume the form

\begin{eqnarray}\label{motion}
\partial_t\Phi&=&p\left[\nabla^2\Phi-\frac{d\mathcal{W}}{d\Phi}-\mathcal{F}
+\gamma(\mathcal{C}-\mathcal{U})\frac{d\mathcal{U}}{d\Phi}\right]\,\,,\nonumber\\
\partial_t\mathcal{C}&=&\nabla\cdot\mathcal{D}\nabla(\mathcal{C}-\mathcal{U})\,\,.
\end{eqnarray}
Here, two independent parameters,

\begin{equation}\label{dimensionless}
\gamma\equiv\frac{\xi\kappa}{\sigma}\left(\frac{\Delta C}{2}\right)^2\,\,\,,\,\,\,p\equiv\frac{V_C}{V_D}\,\,,
\end{equation}
appear, the latter expressed in terms of a crystallization and a diffusion velocity,

\begin{equation}\label{velocities}
V_C\equiv\Gamma\xi\,\,\,,\,\,\,V_D\equiv\frac{D_L}{\xi}\,\,.
\end{equation}

In the following we are going to reduce Eqs. (\ref{motion}) to an interface description, valid up to growth rates of the order of $V_D$, and keeping a finite interface width $2\xi$. This will be done, adopting the specific forms (\ref{doublewell}) and (\ref{drive}) for $\mathcal{W}(\Phi)$ and $\mathcal{F}(\Phi)$, but successively using different choices for the parameters $\gamma$ and $p$, and for the functions $\mathcal{U}(\Phi)$ and $\mathcal{D}(\Phi)$.

\section{Single-Component solidification}

We first consider the case $\gamma=0$ which offers to describe the solidification of the pure solvent by the remainder
\begin{eqnarray}\label{model A}
\frac{1}{p}\,\partial_t\Phi=\nabla^2\Phi-\Theta(-\Phi)(1+F)(\Phi+1)\nonumber\\-\,\Theta(\Phi)(1-F)(\Phi-1)&&
\end{eqnarray}
of Eqs. (\ref{motion}). In this scheme the propagation of a planar solidification front with constant velocity

\begin{equation}\label{V-v}
V=vV_D
\end{equation} 
in $z$-direction, is represented by the single-kink solution

\begin{equation}\label{comoving}
\Phi({\bf r},t)=\Phi_F(z-vt)\equiv\Phi_F(\zeta)
\end{equation}
in the co-moving frame where Eq. (\ref{model A}) reduces to

\begin{eqnarray}\label{steady}
&&\Phi_F''-\Theta(-\zeta)(\Phi_F+1)-\Theta(\zeta)(\Phi_F-1)=\\
&&-\frac{v}{p}\,\Phi_F'+F[\Theta(-\zeta)(\Phi_F+1)-\Theta(\zeta)(\Phi_F-1)]\,\,.\nonumber
\end{eqnarray}
Assuming that $\Phi_F$ is an odd function of $\zeta$, both sides of Eq. (\ref{steady}) vanish separately. As a consequence one finds 

\begin{eqnarray}\label{mov-kink}
\Phi_F(\zeta)&=&\Theta(-\zeta)[-1+\exp{(\zeta)}]\nonumber\\&&
+\Theta(\zeta)[1-\exp{(-\zeta)}]
\end{eqnarray}
which is identical to the equilibrium phase-field profile $\Phi_E(\zeta)$. Moreover, Eq. (\ref{steady}) fixes the relation between the growth rate $v$ and the driving force $F$ in the form

\begin{equation}\label{F-v}
F=\frac{v}{p}\,\,.
\end{equation}
The simple results (\ref{mov-kink}) and (\ref{F-v}) are due to the specific choices (\ref{doublewell}) and(\ref{drive}) of our model. Especially, the former one will allow us to largely copy the derivation of an interface description near thermal equilibrium, presented in Ref. \cite{BDJZ2}. 

In order to study small perturbations of the steady-state solution (\ref{mov-kink}), we linearize Eq. (\ref{model A}) in

\begin{equation}\label{perturbation}
\varphi({\bf x},z-vt,t)\equiv\Phi({\bf r},t)-\Phi_F(z-vt)\,\,.
\end{equation}
This leads to the equation

\begin{equation}\label{linearized}
\left(\frac{1}{p}\,\partial_t-\partial^2+\Omega\right)\varphi({\bf x},\zeta,t)=0\,\,,
\end{equation}
involving the two-dimensional Laplacian 

\begin{equation}\label{laplacian}
\partial^2\equiv\nabla^2-\partial_{\zeta}^2\,\,, 
\end{equation}
and the one-dimensional Schr\"odinger-like operator

\begin{equation}\label{omega}
\Omega\equiv-\partial_{\zeta}^2-F\partial_{\zeta}-2\delta(\zeta)+1+F[\Theta(-\zeta)
-\Theta(\zeta)]\,\,.\end{equation}
By taking the derivative of Eq. (\ref{steady}) with respect to $\zeta$ one finds that $\Phi_F'(\zeta)$ is an eigenfunction of $\Omega$ with zero eigenvalue. It is identical to the bound state of the delta potential in Eq. (\ref{omega}), and also follows from translational symmetry of the system in $\zeta$-direction. The excited state, closest to this ground state, can be found by application of the pseudo-gauge transformation

\begin{eqnarray}\label{gauge}
&&\exp{\left(+\frac{F}{2}\,\zeta\right)}\,\Omega\,\,\exp{\left(-\frac{F}{2}\,\zeta\right)}=\\
&&-\partial_{\zeta}^2-2\,\delta(\zeta)+\left(1+\frac{F}{2}\right)^2\Theta(-\zeta)
+\left(1-\frac{F}{2}\right)^2\Theta(\zeta)\,\,.\nonumber
\end{eqnarray}
The operator (\ref{gauge}) has a band of eigenstates 

\begin{eqnarray}\label{hard-modes}
\psi_k(\zeta)&=&\Theta(-\zeta)\exp{(\kappa\zeta)}\\&&+\Theta(\zeta)\left[\cos{(k\zeta)}
+\frac{\kappa-2}{k}\sin{(k\zeta)}\right]\nonumber
\end{eqnarray}
above the ground state which are parametrized by the wave number $k$. The related eigenvalues read

\begin{equation}\label{eigen-value}
\varepsilon(k)=\left(1+\frac{F}{2}\right)^2-\kappa^2=\left(1-\frac{F}{2}\right)^2+k^2\,\,,
\end{equation}
and are identical to those of the operator $\Omega$. The second identity in Eqs. (\ref{eigen-value}) implies that the band of eigenvalues $\varepsilon(k)$ is separated by a gap $\Delta\,\varepsilon\ge 1/4$ from the ground-state eigenvalue $\varepsilon=0$, provided $F\le 1$. This constraint excludes the regime beyond the ghostly spinodal point of the potential (\ref{double-well}), and, in view of the result (\ref{F-v}), is equivalent to the statement $V\le V_C$.

In Eq. (\ref{linearized}) solutions of the form

\begin{equation}\label{capillary}
\varphi({\bf x},\zeta,t)=\Phi_F'(\zeta)\exp{(i\,{\bf q\cdot x}-\varepsilon t)}
\end{equation}
represent a band of soft modes which have been identified in Ref. \cite{GNW} as Goldstone modes due to broken Euclidean symmetry in space. They are well separated from the hard modes, arising from the excited states (\ref{hard-modes}), and physically are over-damped capillary waves. 

In order to implement the interface position ${\bf R}$ as a collective coordinate in the co-moving frame, we follow Ref. \cite{Zia} where any point ${\bf Q}=({\bf x},\zeta)$ near the interface is represented in the form

\begin{equation}\label{curvilinear}
{\bf Q}={\bf R}({\bf s},t)+u\,{\bf N}({\bf s},t)\,\,.
\end{equation}
Here, ${\bf N}$ is a unit vector, normal to the interface, as shown in Fig. 5, and

\begin{equation}\label{internal}
u=\tilde u({\bf x},\zeta,t)\,\,\,,\,\,\,{\bf s}=\{s^1,s^2\}=\tilde{\bf s}({\bf x},\zeta,t)
\end{equation}
define mappings to a normal coordinate, and to a set of curvilinear coordinates within the interface, respectively. With the notation $\partial_i\equiv\partial/\partial s^i$ the rate of any quantity 

\begin{equation}\label{Psi}
\Psi({\bf s},u,t)\equiv\tilde\Psi({\bf x},\zeta,t)
\end{equation}
at constant ${\bf x}$ and $\zeta$ can be written in the form 

\begin{equation}\label{dot}
\partial_t\tilde\Psi=(\partial_u\Psi)\,\partial_t\tilde u+(\partial_i\Psi)\,\partial_t\tilde s^{\,i}+\partial_t\Psi
\end{equation}

\begin{center}
\begin{figure}
 \includegraphics[width=8cm]{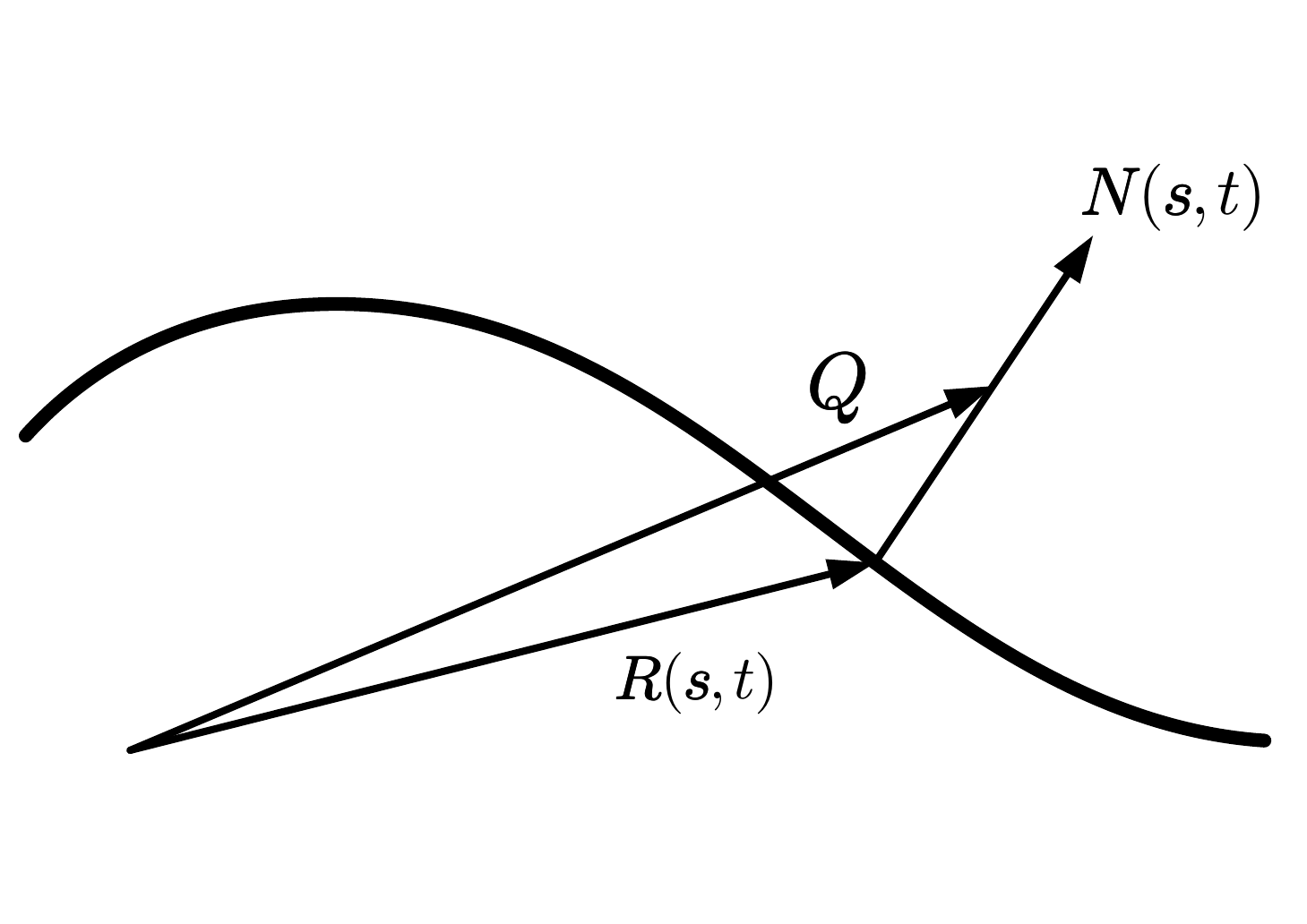}
 \caption{\label{curvi-linear}{Definition of the local curvilinear coordinates $\{{\bf s},u\}$ where ${\bf Q}\equiv\{{\bf x},\zeta\}$ is some position close to the point ${\bf R}({\bf s},t)$ at the interface, and ${\bf N}({\bf s},t)$ is the normal vector at ${\bf R}({\bf s},t)$.}}
 \end{figure}
 \end{center}

The rates $\partial_t\tilde u$ and $\partial_t\tilde s^{\,i}$ are given by the relations

\begin{eqnarray}\label{rates}
&&\partial_t\tilde u=-{\bf N}\cdot \partial_t{\bf R}\,\,,\\
&&(g_{ij}-u\,K_{ij})\,\partial_t\tilde{\,s}^i=-(\partial_t{\bf R}+u\,\partial_t{\bf N})\cdot\partial_j{\bf R}\nonumber
\end{eqnarray}
which follow from Eq. (\ref{curvilinear}), and involve the metric and extrinsic curvature tensors of the interface, 

\begin{eqnarray}\label{metric}
g_{ij}&\equiv&(\partial_i{\bf R})\cdot\partial_j{\bf R}\,\,,\\
K_{ij}&\equiv&-(\partial_i{\bf N})\cdot\partial_j{\bf R}\,\,.\nonumber
\end{eqnarray}
As shown in Ref. \cite{Zia}, the metric tensor of the three-dimensional embedding space, induced by Eq. (\ref{curvilinear}), is block diagonal with $G_{uu}=1$, and with the $2$-dimensional sub-matrix

\begin{equation}\label{Metric}
G_{ij}=(g_{ik}-uK_{ik})\,g^{kl}(g_{lj}-uK_{lj})\,\,,
\end{equation}
where $g_{ij}g^{jk}=\delta_i^k$. This form is needed to evaluate the Laplace-Beltrami representation

\begin{equation}\label{Beltrami}
\nabla^2\tilde\Psi=\frac{1}{\sqrt{G}}\biggl(\partial_i\sqrt{G}G^{ij}\partial_j+\partial_u \sqrt{G}\partial_u\biggr)\Psi\,\,,
\end{equation}
again with $G_{ij}G^{jk}=\delta_i^k$, and $G\equiv Det\{G_{ij}\}$. 

We next decompose $\Phi({\bf r},t)$ in the form

\begin{equation}\label{decomposition}
\Phi({\bf r},t)=\Phi_F(u)+\eta({\bf s},u,t)\,\,,
\end{equation}
along with the Fadeev-Popov condition \cite{FP}

\begin{equation}\label{Fad-Pop}
\int_{-\infty}^{+\infty}du\,\Phi_F'(u)\,\eta({\bf s},u,t)=0\,\,.
\end{equation}
This condition ensures that $\eta({\bf s},u,t)$ is a pure hard-mode field, and implicitly defines ${\bf R}({\bf s},t)$ as a collective field variable. Since excitation of the contribution $\eta$ requires to overcome the gap $\Delta\,\varepsilon$ in the eigenvalue spectrum of the operator (\ref{omega}), $\eta$ may be treated as a perturbation in Eqs. (\ref{model A}). An expansion of this equation to linear order in $\eta$ yields 
 
\begin{equation}\label{linear-order}
\Phi_F'(u)\left(\frac{1}{p}\,\partial_t\tilde u-\nabla^2\tilde u\right)+O(\eta)=0
\end{equation}
where the zeroth-order term derives from Eqs. (\ref{dot}) and (\ref{Beltrami}), applied to $\Psi(u)= \Phi_F(u)$. All other terms of zeroth order cancel due to the identity (\ref{steady}), with $\zeta$ replaced by $u$, and $v$ by the normal velocity component $v_N=vN_z$. If, by multiplication with $\Phi_F'(u)$ and integration over $u$, Eq. (\ref{linear-order}) is projected onto the soft-mode subspace, the terms linear in $\eta$ either drop out due to Eq. (\ref{Fad-Pop}), or they carry pre-factors of the type (\ref{rates}) or (\ref{metric}), and, due to this, are of higher order in our basically hydrodynamic approach.

In view of Eqs. (\ref{rates}) and (\ref{Beltrami}) this procedure leads to the result 

\begin{equation}\label{cap-wave}
\frac{1}{p}\,{\bf N}\cdot\partial_t{\bf R}=\int_{-\infty}^{+\infty}du\,[\Phi_F'(u)]^2\,Tr\left[\frac{{\bf K}}{{\bf g}-u{\bf K}}\right]
\end{equation}
adopting matrix notations for the metric and curvature tensors. The singularity, emerging in the integrand of this expression, if $u$ approaches the smallest curvature radius of the interface, is cured by the factor $[\Phi_F'(u)]^2$ which decays on the scale of the interface thickness. As discussed in Ref. \cite{Zia}, this allows to expand the singular part in Eq. (\ref{cap-wave}) in powers of u, giving rise to moments of the weight $[\Phi_F'(u)]^2$. Following this line, one finds, up to higher-order curvature corrections,

\begin{equation}\label{mean-curvature}
\frac{1}{p}\,{\bf N}\cdot\partial_t{\bf R}=Tr\left[{\bf K}\cdot{\bf g}^{-1}\right]\equiv K
\end{equation}
where $K$ is the local mean curvature of the interface.

We now add the result (\ref{mean-curvature}) as a perturbation to the relation (\ref{F-v}), and, for an under-cooling from the melting temperature $T_M$ to some temperature $T<T_M$, assume

\begin{equation}\label{F-T}
F=L\,\frac{T_M-T}{T_M}\,\,.
\end{equation}
After going back to physical units via Eqs. (\ref{rescaling}), (\ref{dimensionless}), and (\ref{velocities}), we then arrive at the Gibbs-Thomson relation

\begin{equation}\label{Gibbs-Thomson}
L\,\frac{T_M-T}{T_M}=-\sigma K+\frac{\sigma}{D}\frac{V_D}{V_C }(V+{\bf N}\cdot\partial_t{\bf R})\,\,,
\end{equation}
including a kinetic under-cooling term. It locally relates the growth rate to the curvature of the interface under isothermal conditions. We mention that the derivation of this result neither relies on a sharp-interface description nor on a restriction to the low-velocity regime.   

For practical calculations it is convenient, to evaluate Eq. (\ref{mean-curvature}) in a Monge representation with respect to the planar steady-state motion. In this representation the interface-position and normal vectors are given by

\begin{eqnarray}\label{frame}
{\bf R}({\bf s},t)&=&\{{\bf s},h({\bf s},t)\}\,\,,\nonumber\\
{\bf N}({\bf s},t)&=&\frac{1}{\sqrt{g({\bf s},t)}}\{-\partial h({\bf s},t),1\}\,\,.
\end{eqnarray} 
Here, $h({\bf s},t)$ is a local height variable, $\partial\equiv(\partial_1,\partial_2)$ the in-plane nabla operator, and

\begin{equation}\label{Jacobian}
g({\bf s},t)\equiv 1+[\partial h({\bf s},t)]^2 
\end{equation}
is the Jacobian of the interface metric. The metric and extrinsic-curvature tensors read in Monge representation

\begin{equation}\label{Monge}
g_{ij}=\delta_{ij}+(\partial_i h)(\partial_j h)\,\,\,,\,\,\,K_{ij}=\frac{1}{\sqrt{g}}\, \partial_i\partial_j h\,\,. 
\end{equation}
Implementation of these expressions into Eq. (\ref{mean-curvature}) leads to the differential equation

\begin{equation}\label{h-eq}
\frac{1}{\sqrt{g}}\,\partial_t h=p\,\delta^{ij}\,\partial_i\frac{1}{\sqrt{g}}\,\partial_j h\,\,.
\end{equation}
For a stability analysis of the uniform reference motion (\ref{steady}) it is sufficient to linearize this equation in $ h({\bf s},t)$. The result is a two-dimensional diffusion equation with a diffusion constant $p$ which, within this model, implies stability of the planar interface morphology.

In terms of physical units, the nonlinear equation (\ref{h-eq}) can be written in the equivalent form

\begin{equation}\label{Z-eq}
\frac{1}{\sqrt{g}}\,\partial_t h=-\Lambda\frac{\delta H_D}{\delta h}\,\,,
\end{equation}
involving an effective Hamiltonian

\begin{equation}\label{drumhead}
H_D=\frac{\sigma}{2}\int d^2s\,\sqrt{g}\,\,,
\end{equation}
and an Onsager coefficient

\begin{equation}\label{Lambda}
\Lambda\equiv\frac{\Gamma\xi^2}{\sigma}\,\,.
\end{equation}
This representation can be read as a dynamical version of the drumhead model which, complemented by Langevin forces, has been established and analyzed with regard to critical fluctuations in Ref. \cite{BDJZ1}.

\section{Binary-Alloy Solidification}

The inclusion of a finite amount of solute particles is most easily accomplished, if in the model equations (\ref{Hamiltonian}) and (\ref{dynamics}) we choose 

\begin{equation}\label{singlewell}
\mathcal{U}(\Phi)=\frac{C_L+C_S}{2}+\frac{C_L-C_S}{2}\,\Phi\,\,,
\end{equation}
in accordance with the constraints (\ref{C-L,C-S}), and 

\begin{equation}\label{two-sided}
\mathcal{D}(\Phi)=D_L=D_S\equiv D\,\,.
\end{equation}
Adopting this together with the choices (\ref{doublewell}) and (\ref{drive}) for $\mathcal{W}$ and $\mathcal{F}$, the scaled equations of motion (\ref{motion}) assume the explicit form

\begin{eqnarray}\label{linear}
\partial_t\Phi&=&p\left[\nabla^2\Phi-\Theta(-\Phi)(1+F)(\Phi+1)\right.\nonumber\\&&-\,\left. \Theta(\Phi)(1-F)(\Phi-1)+\gamma(\mathcal{C}-1-\Phi)\right]\,\,,\nonumber\\\partial_t\mathcal{C}&=&
\nabla^2(\mathcal{C}-\Phi)
\end{eqnarray}
where, according to Eqs. (\ref{rescaling}), $\mathcal{C}({\bf r},t)$ means the scaled excess concentration with respect to $C_S$.

In the one-dimensional stationary case,  

\begin{eqnarray}\label{steady-state}
&&\Phi({\bf r},t)=\Phi_F(z-vt)\equiv \Phi_F(\zeta)\,\,,\nonumber\\
&&\mathcal{C}({\bf r},t)=C_F(z-vt)\equiv C_F(\zeta)\,\,,
\end{eqnarray}
the last equation in Eqs. (\ref{linear}) can be integrated once. Since, by definition $C_F(-\infty)\equiv 0$, one obtains

\begin{equation}\label{station-c}
C_F'+vC_F=\Phi_F'
\end{equation}
which, due to the behavior $\Phi_F'(+\infty)=C_F'(+\infty)=0$, implies the steady-state boundary condition

\begin{equation}\label{c-bound}
C_F(+\infty)=C_F(-\infty)\equiv 0\,\,.
\end{equation}
In the equation 

\begin{eqnarray}\label{station-Phi}
&&\Phi_F''+\frac{v}{p}\,\Phi_F'-\Theta(-\zeta)(1+F+\gamma)(\Phi_F+1)\nonumber\\
&&-\Theta(\zeta)(1-F+\gamma)(\Phi_F-\Phi_L)=-\gamma C_F\,\,,
\end{eqnarray}
resulting from the first of the Eqs. (\ref{linear}), the condition (\ref{c-bound}) enforces the identification

\begin{equation}\label{phi-bound}
\Phi_F(+\infty)=\frac{1-F-\gamma}{1-F+\gamma}\equiv\Phi_L\,\,.
\end{equation}
The deviation of this expression from the static value $\Phi_E(+\infty)=1$ is negligible, if $\gamma\ll 1-F$, as also pointed out in Ref. \cite{LBT}. 

Elimination of $C_F$ from Eqs. (\ref{station-c}) and (\ref{station-Phi}) leads to the third-order differential equation

\begin{eqnarray}\label{Phi-F}
(\partial_\zeta+v)\biggl[\Phi_F''+\frac{v}{p}\Phi_F'-\Theta(-\zeta)(1+F+\gamma)(\Phi_F+1)
\biggr.&&\nonumber\\-\Theta(\zeta)(1-F+\gamma)(\Phi_F-\Phi_L)\biggl.\biggr]=-\gamma\Phi_F'\,\,\,\,&&
\end{eqnarray}
which implies the matching conditions 

\begin{eqnarray}\label{matching}
&&\Phi_F(-0)=\Phi_F(+0)=0\,\,,\nonumber\\&&\Phi_F'(-0)=\Phi_F'(+0)\,\,,\nonumber\\
&&\Phi_F''(-0)=\Phi_F''(+0)+2\,\,,
\end{eqnarray}
including in the first line the definition $\Phi_F(0)=0$ of the steady-state kink position. Eq. (\ref{Phi-F}) has the single-kink solution

\begin{eqnarray}\label{Phi-solution}
\Phi_F(\zeta)&=&\Theta(-\zeta)[-1+A_-\exp{(\alpha_-\zeta)}]\nonumber\\
&+&\Theta(\zeta)\left[\Phi_L+B_0\exp{(\beta_0\zeta)}\right.\nonumber\\
&&\,\,\,\,\,\,\,\,\,\,\,\,\,\,\,+\left.B_+\exp{(\beta_+\zeta)}\right]
\end{eqnarray}
where $\alpha_-,\beta_0,\beta_+$ are the roots of the cubic equations

\begin{eqnarray}\label{roots}
&&(\alpha+v)\left[\alpha^2+\frac{v}{p}\alpha-(1+F)\right]=\gamma v\,\,,\nonumber\\ &&(\beta+v) \left[\beta^2+\frac{v}{p}\beta-(1-F)\right]=\gamma v
\end{eqnarray}
with $\alpha_->0,\Re(\beta_0)<0,\Re(\beta_-)<0$. The amplitudes in the solution (\ref{Phi-solution}), and the connection between $F$ and $v$ are determined by the relations

\begin{eqnarray}\label{amplitudes}
&&A_--1=B_0+B_++\Phi_L=0\,\,,\nonumber\\
&&A_-\alpha_--B_0\beta_0-B_+\beta_+=0\,\,,\nonumber\\
&&A_-\alpha_-^2-B_0\beta_0^2-B_+\beta_+^2=2\,\,,
\end{eqnarray}
following from Eqs. (\ref{matching}) and (\ref{Phi-solution}).

In the limit $\gamma\rightarrow 0$ the controlling roots of Eqs. (\ref{roots}) approach, due to Eq. (\ref{F-v}), the values

\begin{equation}\label{a-0}
\alpha_-=1\,\,\,,\,\,\,\beta_0=-v\,\,\,,\,\,\,\beta_+=-1\,\,.
\end{equation}
Accordingly, the term $\propto B_0$ in the solution (\ref{Phi-solution}) features an exponential long-distance behavior which elucidates the non-monotonic phase-field profile, found in Ref. \cite{LSO}. More importantly, however, this behavior endangers the suppression effect of the singularity in Eq. (\ref{cap-wave}). A way out of the problem derives from the observation that the amplitude $B_0$ of the dangerous term obeys the relation

\begin{eqnarray}\label{Ao}
&&B_0\,\left[\beta_0^2+\frac{v}{p}\,\beta_0-(1-F)\right]=\\&&\gamma\left(\frac{v}{\alpha_-+v}- B_+\,\frac{v}{\beta_++v}-2\,\frac{1-F}{1-F+\gamma}\right)\nonumber
\end{eqnarray}
which follows from Eqs. (\ref{roots}) and (\ref{amplitudes}). This relationship suggests, and, in fact, forces us to evaluate the model by a perturbation expansion in $\gamma$.

In the following we choose to consider the quantity $v/p=V/V_C$ as another small quantity. This constitutes a noticeable restriction to the growth rates $V$ of materials where $V_C$ obeys an Arrhenius law, like in intermetallic compounds. In dilute metallic alloys, however, particles need not overcome an activation barrier in order to form a crystalline structure. This has been pointed out by Aziz \cite{Aziz}, and is more explicitly demonstrated by the molecular-dynamics simulation of the crystallization of a Lennard-Jones liquid in Ref. \cite{BGJ}. Consequently, in such materials our approximation scheme allows applications to the rapid-growth regime where growth rates can be of the order of the diffusion velocity $V_D$. In view of these features we now will analyze the solidification process of a binary alloy to lowest order of a double-expansion in $\gamma$ and $v/p$.

Starting point of the expansion are the relations

\begin{eqnarray}\label{projection}
&&\frac{1+F+\gamma}{2}-\frac{1-F+\gamma}{2}\,\Phi_L^2=\\&&
\frac{1}{p}\,v\int_{-\infty}^{+\infty}d\zeta\,[\Phi_F'(\zeta)]^2+\gamma\int_{-\infty}^{+\infty} d\zeta\, \Phi_F'(\zeta)C_F(\zeta)\,\,,\nonumber\\&&C_F(\zeta)= \int_{-\infty}^{\zeta}d\zeta'\, \Phi_F'(\zeta')\exp{[v(\zeta'-\zeta)]}\nonumber
\end{eqnarray} 
which directly follow from Eqs. (\ref{station-Phi}) and (\ref{station-c}). Neglecting terms of order $\gamma^2$ and $\gamma v/p$ in the first of these equations allows the replacement $\Phi_F(\zeta)\rightarrow\Phi_E(\zeta)$. This leads to the expressions

\begin{eqnarray}\label{balance}
&&F(v)-G_F(0)=\frac{1}{p}\,v-G_F(v)\,\,,\\&&C_F(\zeta)=\int_{-\infty}^{\zeta} d\zeta'\,\Phi_E'(\zeta')\exp{[v(\zeta'-\zeta)]}\nonumber
\end{eqnarray} 
where, in the first equation, the quantity

\begin{eqnarray}\label{drag}
G_F&\equiv&-\,\gamma\int_{-\infty}^{+\infty}d\zeta\,\Phi_E'(\zeta)\,C_F(\zeta)\nonumber\\
&=&-\,\gamma\,v\int_{-\infty}^{+\infty}d\zeta\,[C_F(\zeta)]^2
\end{eqnarray}
defines the solute drag in line with the definition, used by Hillert in Ref. \cite{Hillert}. On the left-hand side in the first line of Eq. (\ref{balance}) we have used the identity $2\gamma=-\,G_F(0)$. This term derives from the contribution $(\kappa/2)[U(\Phi)]^2$ in the Hamiltonian (\ref{Hamiltonian}) which, due to the expression (\ref{singlewell}) for $U(\Phi)$, favors the solid phase, and, accordingly, acts as an internal driving force in the steady-state growth process. The second line in Eq. (\ref{drag}) follows from Eqs. (\ref{station-c}), and (\ref{c-bound}), and implies that the right-hand side of Eq. (\ref{balance}) is positive.

Insertion of the kink solution (\ref{mov-kink}) into Eqs. (\ref{balance}) and (\ref{drag}) leads to the explicit results

\begin{center}
\begin{figure}
 \includegraphics[width=8cm]{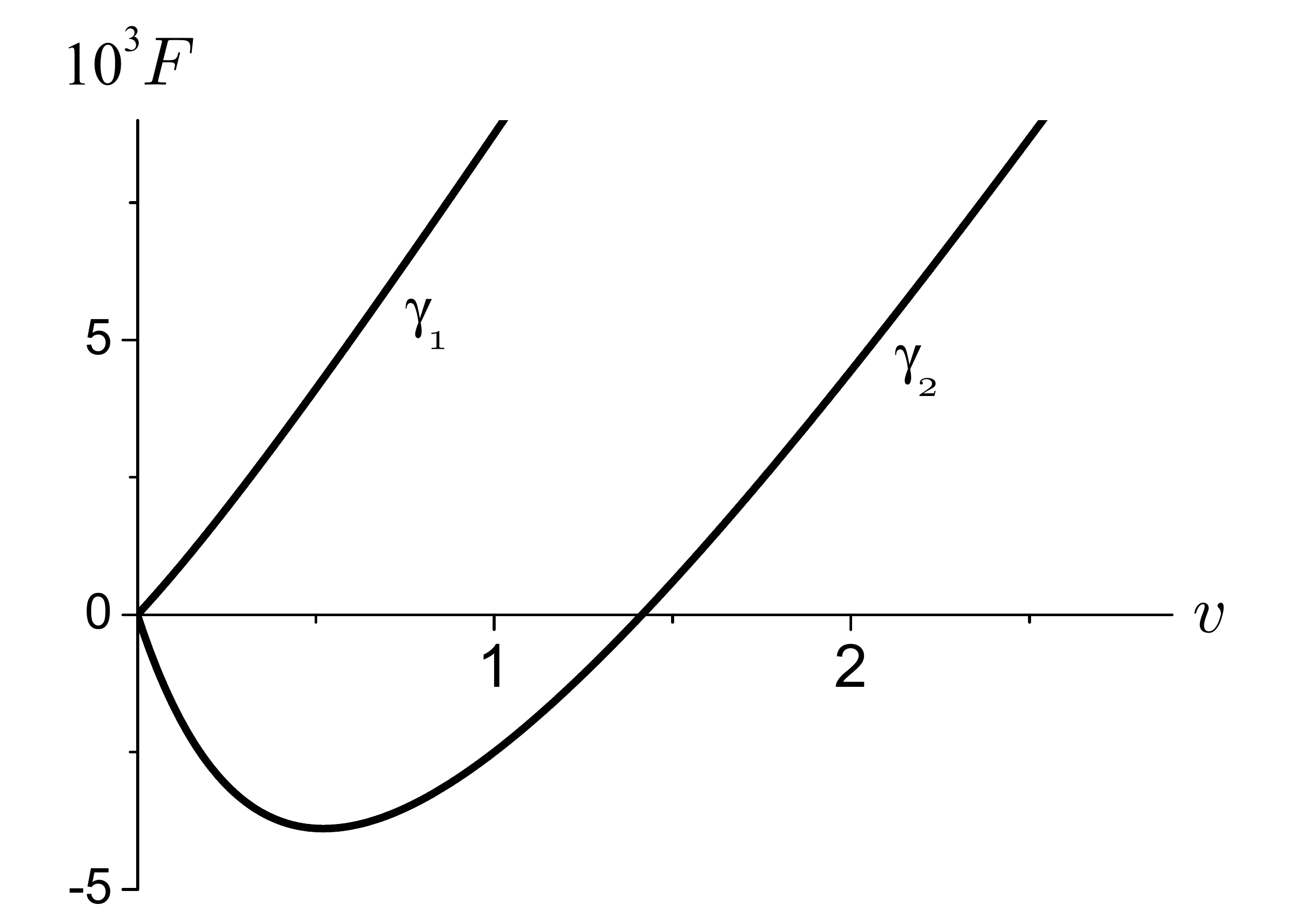}
 \caption{\label{F-anomaly}{The external driving force $F$, considered as a function of the growth rate $v$ according to Eq. (\ref{F,C-F}), for $p=100$, and for $\gamma_1=0.001$ and $\gamma_2=0.01$, respectively. For the latter value $F(v)$ has a minimum, to the left of which the anomaly $F'(v)<0$ is visible.}}
 \end{figure}
 \end{center}

\begin{center}
\begin{figure}
 \includegraphics[width=8cm]{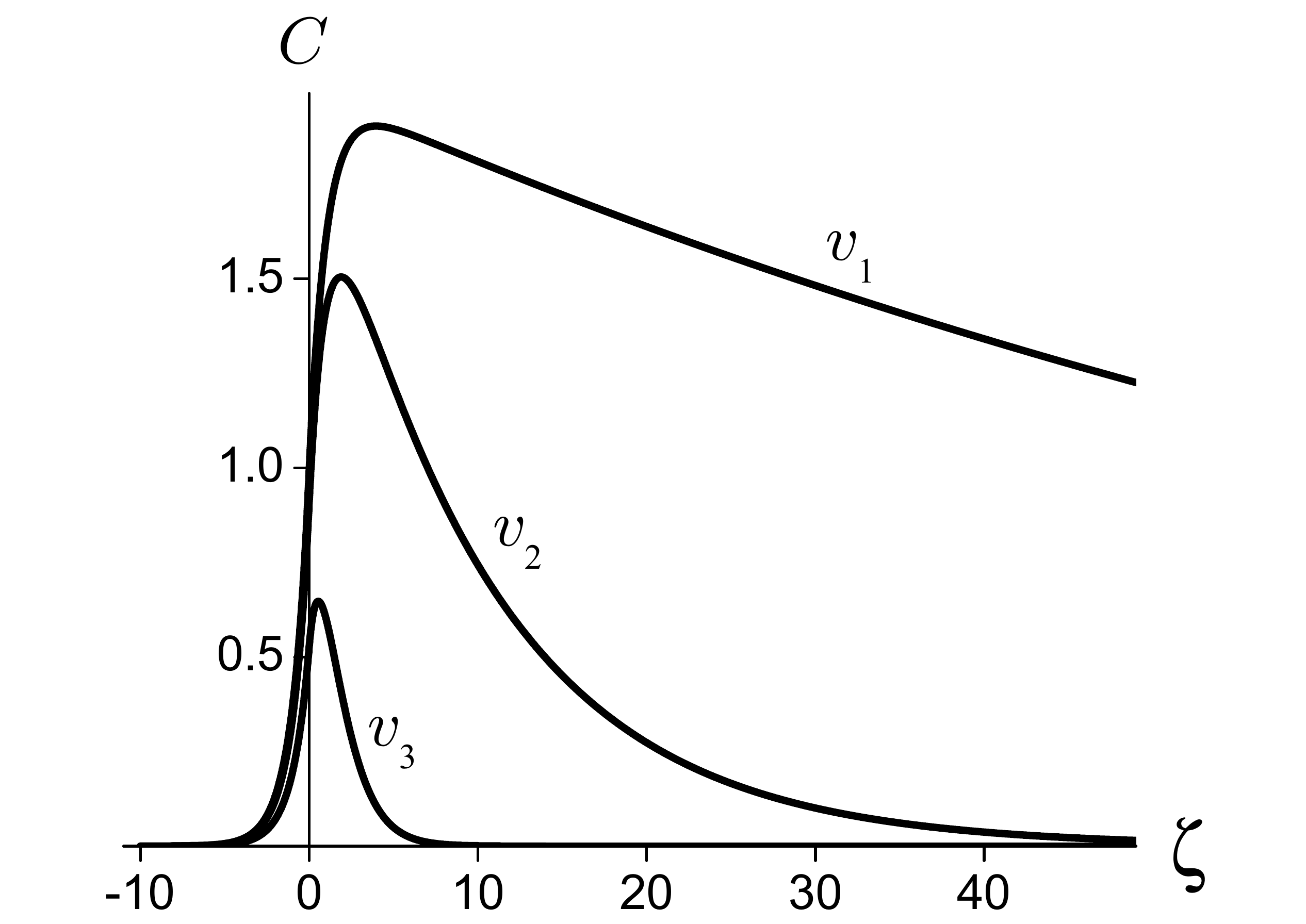}
 \caption{\label{C-profile}{The steady-state concentration profile for the velocity values $v_1=0.001$, $v_2=0.01$, and $v_3=0.9$, illustrating the solute-trapping effect.}}
 \end{figure}
 \end{center}

\begin{eqnarray}\label{F,C-F}
&&F(v)=\frac{v}{p}+\gamma\left[\frac{v+2}{(v+1)^2}-2\right]\,\,,\\ &&C_F(\zeta)=\Theta(-\zeta)\frac{1}{v+1}\exp{(\zeta)}\nonumber\\
&&+\,\Theta(\zeta)\left[\frac{1}{v-1}\exp{(-\zeta)}-\frac{2}{v^2-1}\exp{(-v\zeta)}\right]\,\,.\nonumber
\end{eqnarray}
The first of these equations implies

\begin{equation}\label{anomal}
F'(0)=\frac{1}{p}-\frac{1}{p_c}\,\,\,,\,\,\,p_c\equiv\frac{1}{3\gamma}\,\,.
\end{equation}
Accordingly, $F(v)$ has a positive slope at the origin for $p<p_c$, but shows the anomalous behavior $F'(0)<0$ for $p>p_c$, pointed out already in Ref. \cite{LBT}. In the literature this anomaly has repeatedly been encountered again, however, without any clarification of its physical background. Below, we will demonstrate that the effect is due to a strong instability of the system, suggesting that the inclusion of random forces in the basic equations of motion (\ref{dynamics}) might become important if $F'(v)< 0$. The results (\ref{F,C-F}) are illustrated in Figs. 6 and 7 in the unstable regime $p>p_c$. We mention that the existence of an anomaly of the type $F'(v)< 0$ is also in line with the velocity-dependent interface temperature, obtained in the approach by Aziz and Boettinger \cite{AB}.

The results (\ref{F,C-F}) also determine the partition coefficient

\begin{equation}\label{partition}
K(V)\equiv\frac{C_S}{C_S+C_F(0)\Delta C}
\end{equation}
which measures the solute-trapping effect, and for $V=0$ reduces to the equilibrium value

\begin{equation}\label{E-partition}
K_E\equiv K(0)=\frac{C_S}{C_L}\,\,.
\end{equation}
By insertion of the value $C_F(0)=1/(v+1)$, following from Eq. (\ref{F,C-F}), one recovers the form 

\begin{equation}\label{Aziz}
K(V)=\frac{K_E+V/V_D^*}{1+V/V_D^*}\,\,,
\end{equation}
suggested by Aziz in Ref. \cite{Aziz}. Remarkably, however, we moreover find a reference velocity

\begin{equation}\label{V-D*}
V_D^*\equiv\frac{V_D}{K_E} 
\end{equation}
which shows a dependence on $K_E$ of the type proposed in Ref. \cite{SA}.

In order to examine the stability of the steady-state solutions $\Phi_F,C_F$, we next expand the exact equations of motion (\ref{linear}) to linear order in the perturbations

\begin{eqnarray}\label{deviations}
\varphi({\bf x},z-vt,t)&=&\Phi({\bf r},t)-\Phi_F(z-vt)\,\,,\nonumber\\n({\bf x},z-vt,t)&=&\mathcal{C}({\bf r},t)-C_F(z-vt)\,\,.
\end{eqnarray}
The resulting equations can be written in the form

\begin{eqnarray}\label{matrix}
\left( \begin{array}{ccc}
\partial_t-p\,\partial^2 & 0 \\\partial^2 & \partial_t-\partial^2 \\
\end{array} \right)
\left( \begin{array}{ccc}
\varphi \\n \\
\end{array} \right)&&\nonumber\\
=-\left( \begin{array}{ccc}
p(\Omega+\gamma) & -p\gamma \\\partial_\zeta^2 & -\partial_\zeta^2 -v\partial_\zeta \\
\end{array} \right)
\left( \begin{array}{ccc}
\varphi \\n \\
\end{array} \right)&&
\end{eqnarray}
where, replacing Eq. (\ref{omega}), $\Omega$ now is defined by

\begin{eqnarray}\label{Omega}
\Omega&\equiv&-\partial_{\zeta}^2-\frac{v}{p}\partial_{\zeta}-2\frac{1}{\Phi_F'(0)}\delta(\zeta)
+1\nonumber\\&\,\,\,&+\,F[\Theta(-\zeta)-\Theta(\zeta)]\,\,.
\end{eqnarray}
By taking first and second derivatives of Eqs. (\ref{station-Phi}) and (\ref{station-c}) with respect to $\zeta$, one finds that the matrix operator on the right-hand side of Eq. (\ref{matrix}) obeys the relation

\begin{equation}\label{zero-mode}
\left( \begin{array}{ccc}
p(\Omega+\gamma) & -p\gamma \\\partial_\zeta^2 & -\partial_\zeta^2 -v\partial_\zeta \\
\end{array} \right)
\left( \begin{array}{ccc}
\Phi_F' \\C_F' \\
\end{array} \right)=0
\end{equation}
which means that translational symmetry in $\zeta$-direction again provides an eigenstate with eigenvalue zero.

A convenient way, to explore the appearance of other eigenstates, is to expand the elements of the matrix in Eq. (\ref{zero-mode}) in the small parameters $\gamma$ and $v/p$. To leading order the resulting eigenvalue equation reads

\begin{equation}\label{lead-matr}
\left( \begin{array}{ccc}
p\,\Omega_E &0 \\\partial_\zeta^2 & -\partial_\zeta^2 -v\partial_\zeta \\
\end{array} \right)
\left( \begin{array}{ccc}
\psi \\\theta \\
\end{array} \right)=
p\,\varepsilon\left( \begin{array}{ccc}
\psi \\\theta \\
\end{array} \right)
\end{equation}
where the operator $\Omega_E$ is given by

\begin{equation}\label{hard-mode}
\Omega_E\equiv-\partial_\zeta^2+1-2\delta(\zeta)\,\,.
\end{equation}
The upper component of Eq. (\ref{lead-matr}) yields the autonomous eigenvalue equation 

\begin{equation}\label{Schroedinger}
\Omega_E\,\psi(\zeta)=\varepsilon\,\psi(\zeta)
\end{equation}
whereas the less interesting lower component in principle allows to calculate $\theta(\zeta)$. Eq. (\ref{Schroedinger}) possesses two classes of solutions, corresponding to the ground and scattering states of the operator (\ref{hard-mode}). The ground-state equation is a relict of Eq. (\ref{zero-mode}), implying $\varepsilon=0$ and $\psi(\zeta)=\Phi_E'(\zeta)$. The scattering states form a band with 

\begin{eqnarray}\label{hard-band}
&&\varepsilon(k)=1+k^2\,\,,\\
&&\psi_k(\zeta)\propto\left[\exp{(ik\zeta)}-\frac{1}{1+\vert k\zeta\vert}\exp{(i\vert k\zeta\vert)}\right]\,\,.\nonumber\\
\end{eqnarray}
Again we have the situation that the soft modes of the system, deriving from Eqs. (\ref{matrix}) and (\ref{zero-mode}), are separated from the hard modes by a gap which, in leading order of our expansion, is given by $\Delta\varepsilon=1$. Since this value will only slightly be shifted in our expansion, we will eliminate the hard modes of the system by suitably extending the procedure, described below Eq. (\ref{decomposition}).

As a first step we complement the decomposition (\ref{decomposition}) by splitting $\mathcal{C}$ into three contributions,

\begin{eqnarray}\label{separation}
\mathcal{C}({\bf r},t)&=&C_F(u)+c({\bf s},u,t)+\vartheta({\bf s},u,t)\nonumber\\&\equiv&C_F(u)+\tilde c({\bf x},\zeta,t)+\tilde\vartheta({\bf x},\zeta,t)\,\,.
\end{eqnarray}
Here, by definition, $\vartheta$ is related to 

\begin{equation}\label{eta}
\eta({\bf s},u,t)\equiv\tilde\eta({\bf x},\zeta,t)
\end{equation}
by the equation

\begin{equation}\label{hard-cmodes}
(\partial_t-v\partial_\zeta-\nabla^2)\,\tilde\vartheta\equiv-\nabla^2\tilde\eta
\end{equation}
which is a copy of the second line in Eq. (\ref{linear}), written in the co-moving frame. In view of Eq. (\ref{station-c}) the remaining equation for $\tilde c$ reads
\begin{equation}\label{c-equation}
(\partial_t-v\partial_\zeta-\nabla^2)\,\tilde c=-(\partial_t\tilde u-\nabla^2\tilde u)\,C_F'-(\nabla^2\tilde u)\Phi_E'
\end{equation}
where we have, in the spirit of our expansion scheme, replaced $\Phi_F'$ by $\Phi_E'$.
When the solution $\tilde c$ of this equation is introduced into the first line of Eq. (\ref{linear}), projection onto the soft-mode component $\Phi_E'(u)$ leads to the result

\begin{equation}\label{R-g-eqn}
\frac{1}{p}\,{\bf N}\cdot D_t{\bf R}=K-\textmd{g}
\end{equation}
where the last term appears as a perturbation

\begin{equation}\label{friction}
\textmd{g}({\bf s},t)\equiv-\,\gamma\int_{-\infty}^{+\infty}du\,\Phi_E'(u)\,c(u,{\bf s},t)
\end{equation}
of the drag force (\ref{drag}). Here, in addition to the neglected higher-order terms, leading to Eq. (\ref{mean-curvature}), we have omitted terms of order $\gamma\eta$. The singularities, arising from the terms $\nabla^2\tilde u=-Tr[{\bf K}/({\bf g}-u{\bf K})]$ in Eq. (\ref{c-equation}), are cured by the factor $\Phi_E'(u)$ in Eq. (\ref{friction}). This justifies to replace these terms by $-K$, just as in Eq. (\ref{mean-curvature}). As a result, the equation (\ref{c-equation}) can be written in the form

\begin{equation}\label{final-c}
(\partial_t-v\partial_\zeta-\nabla^2)\,\tilde c=({\bf N}\cdot D_t{\bf R}-K)\,C_F'+K\Phi_E'
\end{equation}
where the differential operator on the left-hand side can be represented in terms of curvilinear coordinates via Eqs. (\ref{dot}) - (\ref{Beltrami}).

Eqs. (\ref{R-g-eqn}) - (\ref{final-c}) are the main results of the present section. Parallel to the procedure, leading to the Gibbs-Thomson relation (\ref{Gibbs-Thomson}), we are going to add the result (\ref{R-g-eqn}) as a perturbation to the force balance in the first line of Eq. (\ref{balance}). For an under-cooling from an initial point $T_L(C_S)$ at the liquidus line to some temperature $T<T_L(C_S)$ at constant concentration we assume

\begin{eqnarray}\label{Undercooling}
F&=&4\gamma\left[\frac{T_L(C_S)-T}{T_L(C_S)-T_S(C_S)}-1\right]\nonumber\\
&=&4\gamma\,\frac{T_S(C_S)-T}{T_L(C_S)-T_S(C_S)}\,\,.
\end{eqnarray}
Here, the pre-factor $4\gamma$ has been chosen such that, after transforming to physical units, we obtain

\begin{eqnarray}\label{noneq.G-T}
L\,\frac{T_S(C_S)-T}{T_M}&=&G_F(0)-G({\bf s},t)-\sigma K\\&+&\frac{\sigma}{D}\frac{V_D}{V_C} (V+{\bf N}\cdot\partial_t{\bf R})\nonumber
\end{eqnarray}
which in a natural way generalizes the previous form (\ref{Gibbs-Thomson}) of the Gibbs-Thomson relation. The new term, involving $G_F(0)$ and the total drag force

\begin{equation}\label{total drag}
G({\bf s},t)\equiv G_F+\textmd{g}({\bf s},t)\,\,,
\end{equation}
acts as an additional under-cooling. A similar effect is present in the result for the interface temperature in the Aziz-Boettinger approach \cite{AB}. We mention that Eq. (\ref{noneq.G-T}) also provides a new answer to the question in Ref. \cite{Davis}, concerning a Gibbs-Thomson equation, valid for non-planar interfaces.
 
The rather involved evaluation of Eqs. (\ref{R-g-eqn}) - (\ref{final-c}) is considerably simplified, if the  Monge representation of these equations is only used in linear order in $h$ and $c$. This is, fortunately, adequate for a stability analysis of the planar morphology of the interface, and leads to the differential equations

\begin{eqnarray}\label{h-c-eqs}
&&\partial_t h=p\left[\partial^2 h+\textmd{g}\right]\,\,,\\
&&(\partial_t-v\partial_\zeta-\partial_\zeta^2-\partial^2)\,c=
C_F'(\partial_t-\partial^2)h+\Phi_E'\,\partial^2 h\nonumber
\end{eqnarray}
where $\textmd{g}$ is given by Eq. (\ref{friction}). Within this approximation the metric tensor in Eq. (\ref{Monge}) reduces to that of a planar geometry which allows us to identify the set of curvilinear coordinates ${\bf s}$ with the set ${\bf x}$ of Euclidean coordinates. In order to finally transform the equations (\ref{h-c-eqs}) to the laboratory frame and to physical units, we define the interface position

\begin{equation}\label{Z}
Z({\bf x},t)\equiv Vt+h({\bf x},t)\,\,,
\end{equation}
and the soft-mode concentration field

\begin{eqnarray}\label{C}
C({\bf r},t)&\equiv&\mathcal{C}({\bf r},t)-\vartheta[{\bf x},z-Z({\bf x},t),t]\\
&=&C_F[z-Z({\bf x},t)]+c({\bf x},z-Vt,t)+O(h^2)\,\,.\nonumber
\end{eqnarray}
Assembling the results (\ref{balance}), (\ref{drag}), and (\ref{h-c-eqs}), written in physical units, we obtain, remembering the definitions (\ref{Lambda}), and (\ref{singlewell}), the set of equations

\begin{eqnarray}\label{Z-C-dynamics}
&&\partial_t Z=\Lambda\biggl\{F+\sigma\,\partial^2 Z\biggr.\nonumber\\&&-\left.\kappa\int_{-\infty}^{+\infty}
dz\Bigl(C-U\bigl[\Phi_E(z-Z)\bigr]\Bigr)\partial_zU\bigl[\Phi_E(z-Z)\bigr]\right\}\,\,,\nonumber\\
&&\partial_t C=D\,\nabla^2\Bigl\{C-U\bigl[\Phi_E(z-Z)\bigr]\Bigr\}
\end{eqnarray}
which determines the uniform propagation of the planar solidification front, including its stability behavior.

Eqs. (\ref{Z-C-dynamics}) present the requested interface description of the solidification process which generally holds at low growth velocities, but also applies to the rapid-growth regime of dilute metallic alloys. It, moreover, reconciles with the presence of a finite interface width, entering via the static phase-field profile $\Phi_E$. Remarkably, the latter is the only remnant of the initial phase-field description. This feature is a consequence of our expansion scheme which has been established by studying the phase-field model with the specific choices (\ref{singlewell}) and (\ref{two-sided}). Once, this scheme has been accepted, the interface model essentially follows by projecting the phase-field equation of motion to the soft-mode component $\Phi_E'$. Application of the same procedure to the more general phase-field model, given  by Eqs. (\ref{Hamiltonian}), (\ref{dynamics}), and (\ref{drive}), leads to an interface version where the quantities $\Phi_E(z-Z)$ and $D$ are replaced in proper positions by the functions 

\begin{eqnarray}\label{U-D}
U(z-Z)&\equiv&\mathcal{U}[\Phi_E(z-Z)]\,\,,\nonumber\\
D(z-Z)&\equiv&\mathcal{D}[\Phi_E(z-Z)]\,\,.
\end{eqnarray}
Similar input functions enter the description of grain-boundary motion by Cahn \cite{Cahn} who, however, uses an ideal-gas picture of the impurities in a Fokker-Planck representation of the system. In the following we are going to discuss some properties and applications of our generalized interface model.
 
\section{Capillary-Wave Description}

Our first observation is that the new interface model can be written in terms of an effective Hamiltonian

\begin{eqnarray}\label{I-Hamiltonian}
H&=&\frac{\sigma}{2}\int d^2x\,(\partial Z)^2\nonumber\\
&+&\frac{\kappa}{2}\int d^3r\,\Bigl[C-U(z-Z)\Bigr]^2\,\,.
\end{eqnarray}
Even without recourse to the phase-field description it is clear from the equilibrium condition $\delta H/\delta C=0$ that 

\begin{equation}\label{C_E-U}
U(z-Z)=C_E(z-Z)
\end{equation}
is identical to the equilibrium concentration profile of the solute component. The equations of motion for the field variables $Z({\bf x},t)$ and $C({\bf r},t)$ are obtained in the form

\begin{eqnarray}\label{I-dynamics}
\partial_t Z&=&\Lambda\,\left(F-\frac{\delta H}{\delta Z}\right)\,\,,\nonumber\\
\partial_t C&=&\nabla\cdot D(z-Z)\,\nabla\,\frac{1}{\kappa}\,\frac{\delta H}{\delta C}\,\,,
\end{eqnarray}
which can again be justified without going back to the phase-field description. In fact, in thermal equilibrium, where, instead of the driving force $F$, conveniently chosen Langevin forces enter \cite{BDJZ2}, the form of the equations (\ref{I-dynamics}) essentially follows from the principle of detailed balance. As another aspect, we mention that, after insertion of the Hamiltonian (\ref{I-Hamiltonian}), the equations of motion (\ref{I-dynamics}) are of a hydrodynamic type where capillary waves are, in the simplest-possible way, coupled to a bulk-diffusion field. 

Within this approach the uniform motion of a planar solidification front is, in dimensionless form, described by the unchanged first equation in Eqs. (\ref{balance}). However, the drag force is now determined by the expressions

\begin{eqnarray}\label{I-G-C}
&&G_F(v)=-\,\gamma\int_{-\infty}^{\infty}d\zeta\,U'(\zeta)\,C_F(\zeta)\,\,,\\
&&C_F(\zeta)=\int_{-\infty}^{\zeta}d\zeta'\,U'(\zeta')\,\exp{-\,\int_{\zeta'}^{\zeta}
d\zeta''\frac{v}{D(\zeta'')}}\nonumber
\end{eqnarray}
where $D(z-Z)$ has been re-scaled according to the last of the Eqs. (\ref{rescaling}). The expression for $C_F(\zeta)$ in Eqs. (\ref{I-G-C}) derives from the once-integrated equation of motion

\begin{equation}\label{C_F-U}
C_F'(\zeta)+\frac{v}{D(\zeta)}C_F=U'(\zeta)
\end{equation}
which for the drag force $G_F(v)$ implies the representation

\begin{equation}\label{drag-U}
G_F(v)=-\,\gamma\int_{-\infty}^{\infty}d\zeta\,\frac{v}{D(\zeta)}\,[C_F(\zeta)]^2\,\,.
\end{equation}
From Eqs. (\ref{I-G-C}) and (\ref{drag-U}) we conclude that generally

\begin{equation}\label{G-properties}
G_F(0)=-2\gamma\,\,\,,\,\,\,G_F(\infty)=0\,\,\,,\,\,\,G_F(v)\le 0
\end{equation}
where in the first equality we have used the behavior $U(+\infty)-U(-\infty)=2$, following after re-scaling from Eqs. (\ref{C-L,C-S}). Explicit evaluations of the quantities (\ref{I-G-C}), of course, require specific choices of the functions $U(z-Z)$ and $D(z-Z)$.

For models, characterized by an arbitrary equilibrium-concentration profile $U(z-Z)$, but a uniform diffusion coefficient $D(z-Z)=1$, it is possible to determine all unstable eigenmodes of the solidification front. In order to demonstrate this, we start from the equations (\ref{friction}) and (\ref{h-c-eqs}) with $\Phi_E'(\zeta)$ replaced by $U'(\zeta)$. If, moreover, we consider perturbations of the form

\begin{eqnarray}\label{monochrom}
&&h({\bf x},t)=\hat h({\bf q},\omega)\exp{(i{\bf q}\cdot{\bf x}+\omega t)}\,\,,\\
&&c(\zeta,{\bf x},t)=\hat c(\zeta,{\bf q},\omega)\exp{(i{\bf q}\cdot{\bf x}+\omega t)}\,\,,\nonumber
\end{eqnarray}
we encounter the set of equations

\begin{eqnarray}\label{h-g-c}
&&(\omega+p\,q^2)\hat h({\bf q},\omega)=-\,p\,\gamma\int_{-\infty}^{+\infty}du\,U'(\zeta)\,\hat c(\zeta,{\bf q},\omega)\,\,,\nonumber\\
&&(\omega+q^2-v\partial_\zeta-\partial_\zeta^2)\,\hat c(\zeta,{\bf q},\omega)=\nonumber\\
&&C_F'(\zeta)\,(\omega+q^2)\hat h({\bf q},\omega)-U'(\zeta)\,q^2\hat h({\bf q},\omega)\,\,.
\end{eqnarray}
The last equation is a differential equation for $\hat c$ which, by the substitution

\begin{equation}\label{substitution}
\rho(\zeta,{\bf q},\omega)\equiv\frac{\hat c(\zeta,{\bf q},\omega)}{\hat h({\bf q},\omega)}-C_F'(\zeta)
\end{equation}
can be converted into the more convenient form

\begin{eqnarray}\label{rho-eqn}
&&(\omega+q^2-v\partial_\zeta-\partial_\zeta^2)\,\rho(\zeta,{\bf q},\omega)=\nonumber\\
&&U'''(\zeta)-q^2U'(\zeta)
\end{eqnarray}
where the source term on the right-hand side is directly expressed in terms of the input function $U(\zeta)$. Eq. (\ref{rho-eqn}) has the solution

\begin{eqnarray}\label{rho-solution}
\rho(\zeta,{\bf q},\omega)&=&-\,U'(\zeta)\biggr.\\&+&\frac{\lambda^2-q^2}{\lambda-\mu}\int_\zeta^\infty d\zeta'\,U'(\zeta')\exp{[\lambda(\zeta-\zeta')]}\nonumber\\&+& \frac{\mu^2-q^2}{\lambda-\mu}\int_{-\infty}^\zeta d\zeta'\,
U'(\zeta')\exp{[\mu(\zeta-\zeta')]}\nonumber
\end{eqnarray}
with the characteristic roots $\lambda,\mu$ given by

\begin{equation}\label{kappa}
\lambda\equiv -\frac{v}{2}+\sqrt{\frac{v^2}{4}+\omega+q^2}\equiv -\mu-v\,\,.
\end{equation}
Insertion of this solution into the first line of Eqs. (\ref{h-g-c}) leads to the eigenvalue equation 

\begin{equation}\label{homogeneous}
[\omega+p\,q^2+\Sigma({\bf q},\omega)]\hat h({\bf q},\omega)=0 
\end{equation}
where we have introduced a kind of self energy,

\begin{eqnarray}\label{self-energy}
\Sigma({\bf q},\omega)&\equiv&\,p\,\gamma\int_{-\infty}^{+\infty}du\,U'(\zeta)\,[\rho(\zeta,{\bf q},\omega)+C_F'(\zeta)]\nonumber\\&=&-p\,v[G_F(v+\lambda)-G_F(v)]\nonumber\\
&\,&-p\,\frac{q^2-\lambda^2}{v+2\lambda}\,[G_F(v+\lambda)+G_F(\lambda)]\,\,.
\end{eqnarray}
The final expression for $\Sigma({\bf q},\omega)$ has been obtained from Eqs. (\ref{station-c}) and (\ref{rho-solution}), and is identical to that found in Ref. \cite{PRL} for the impurity-controlled motion of general domain boundaries.

According to Eq. (\ref{self-energy}), non-trivial solutions $\hat h({\bf q},\omega)$ of Eq. (\ref{homogeneous}) only exist under the condition 

\begin{eqnarray}\label{dispersion}
&&\omega+p\,q^2-p\,v[G_F(v+\lambda)-G_F(v)]\nonumber\\
&&-p\,\frac{\lambda^2-q^2}{v+2\lambda}\,[G_F(v+\lambda)+G_F(\lambda)]=0\,\,.
\end{eqnarray}
This relation determines the amplification rates $\omega(q)$ of all unstable eigenmodes of the model. It has a universal character, since it applies to a whole family of models with a globally uniform diffusion constant $D$, but with different equilibrium concentration profiles $U(\zeta)$. Given a specific form of such a profile, the explicit evaluation of Eq. (\ref{dispersion}) only requires the knowledge of $G_F(v)$ which follows from an analysis of the related one-dimensional growth scenario. 

By an expansion of the dispersion relation (\ref{dispersion}) in $q$ and $\omega$ we find two branches $\omega_1(q)$ and $\omega_2(q)$ with the behavior $\omega_1(0)=\omega_1'(0)=\omega_2'(0)=0$, and

\begin{eqnarray}\label{two-modes}
\omega_1''(0)&=&2\,\frac{G_F'(v)-[G_F(v)+G_F(0)]/v-1}{1/p-G_F'(v)}\,\,,\nonumber\\
\omega_2(0)&=&2\,\frac{\,1/p-G_F'(v)\,}{\{[G_F(v)+G_F(0)]/v\}''}\,\,.
\end{eqnarray}
These expressions have a universal form in the same sense as Eq. (\ref{dispersion}). As demonstrated in Ref. \cite{PRB}, they even apply to models where the diffusion coefficient $D(z-Z)$ has a constant value $D$ in the liquid and the interface region, but is zero in the solid phase. Since for $v\rightarrow 0$ the numerator in $\omega_1''(0)$ shows the behavior $2\vert G_F(0)\vert/v$, but approaches the value $-1$ for $v\rightarrow\infty$, the mode $\omega_1(q)$ is unstable at low and stable at large velocities, provided the denominator in $\omega_1''(0)$ is positive. Due to the first line in Eq. (\ref{balance}) this denominator is
given by $F'(v)$ which, as demonstrated below Eq. (\ref{anomal}), can become negative. In this case the second mode $\omega_2(q)$ becomes unstable, since the denominator in $\omega_2(0)$ turns out to be negative in all applications of interest. An unstable point of the type $q=0,\,\omega_2(0)>0$ has previously been discovered by Cahn in a one-dimensional model of impurity-controlled grain-boundary motion \cite{Cahn}, in view of which we will denote the instability of the mode $\omega_2(q)$ as a Cahn instability.

If, in a first application, we reconsider the model with 

\begin{equation}\label{U-exponential}
U(\zeta)=\Phi_E(\zeta)+1\,\,\,,\,\,\,D(\zeta)=1\,\,,
\end{equation}
we find, in accordance with the first line in Eqs. (\ref{F,C-F}),

\begin{equation}\label{G-exponential}
G_F(v)=-\,\gamma\,\frac{v+2}{(v+1)^2}\,\,.
\end{equation}

\begin{center}
\begin{figure}
 \includegraphics[width=8cm]{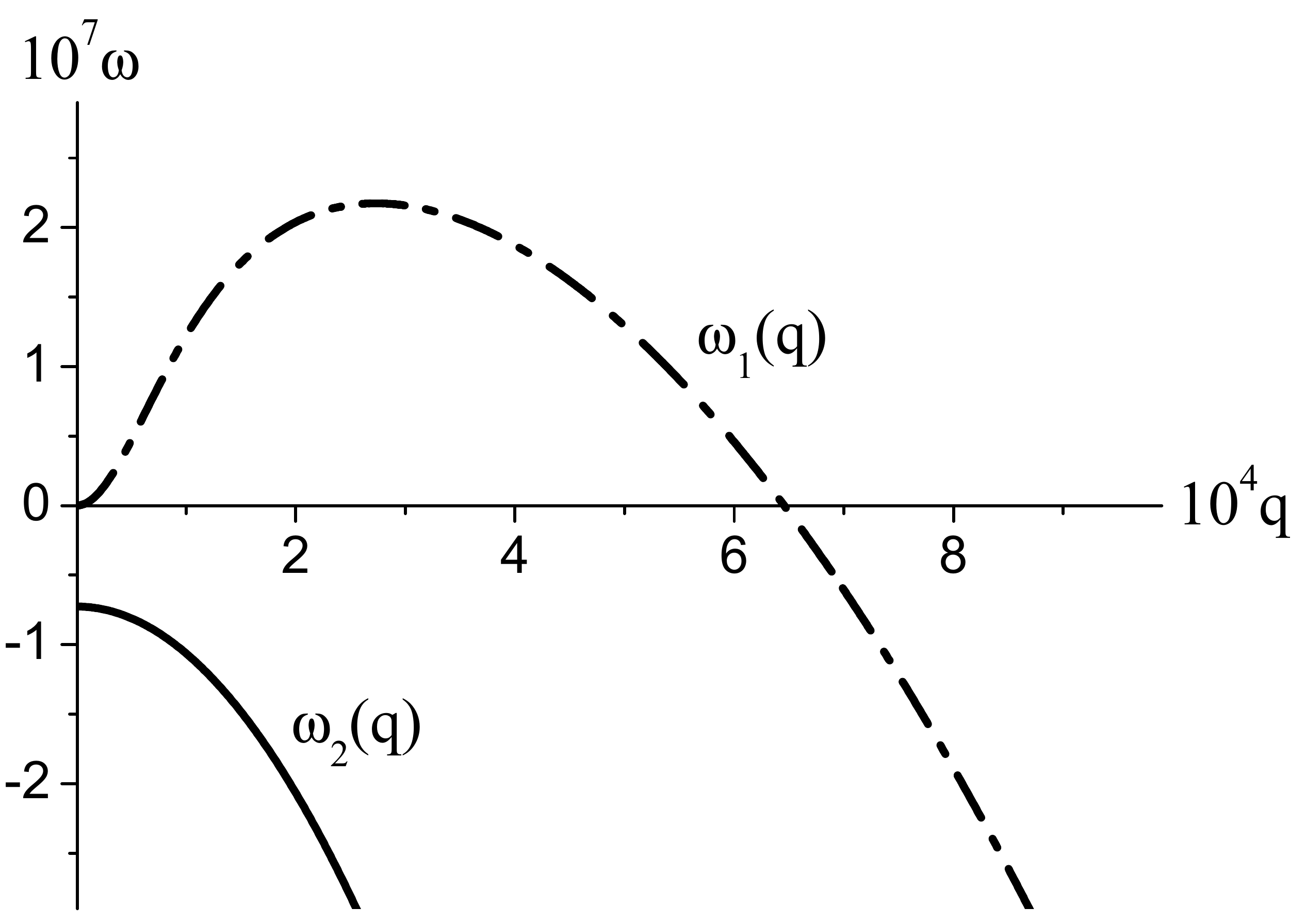}
 \caption{\label{MS-instability}{Dispersion curve of the unstable Mullins-Sekerka-like mode $\omega_1(q)$, and of the stable mode $\omega_2(q)$ for the parameter values $p=100$, $\gamma=0.001$, and $v=0.0036$.}}
 \end{figure}
 \end{center}

\begin{center}
\begin{figure}
 \includegraphics[width=8cm]{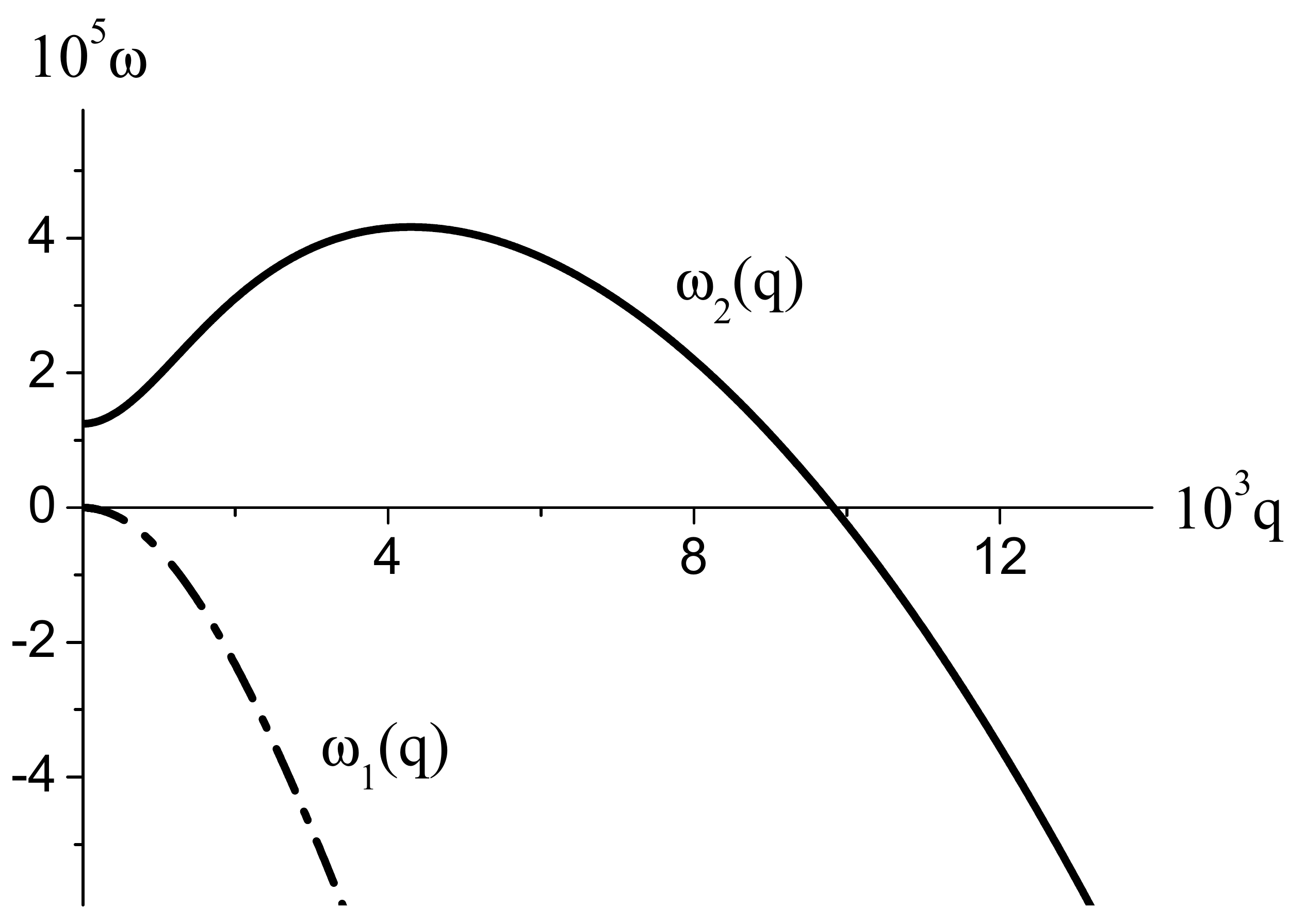}
 \caption{\label{Cahn-instability}{Dispersion curve of the unstable Cahn mode $\omega_2(q)$, and of the stable mode $\omega_1(q)$ for the parameter values $p=100$, $\gamma=0.01$, $v=0.03$.}}
 \end{figure}
 \end{center}
 
\begin{center}
\begin{figure}
 \includegraphics[width=8cm]{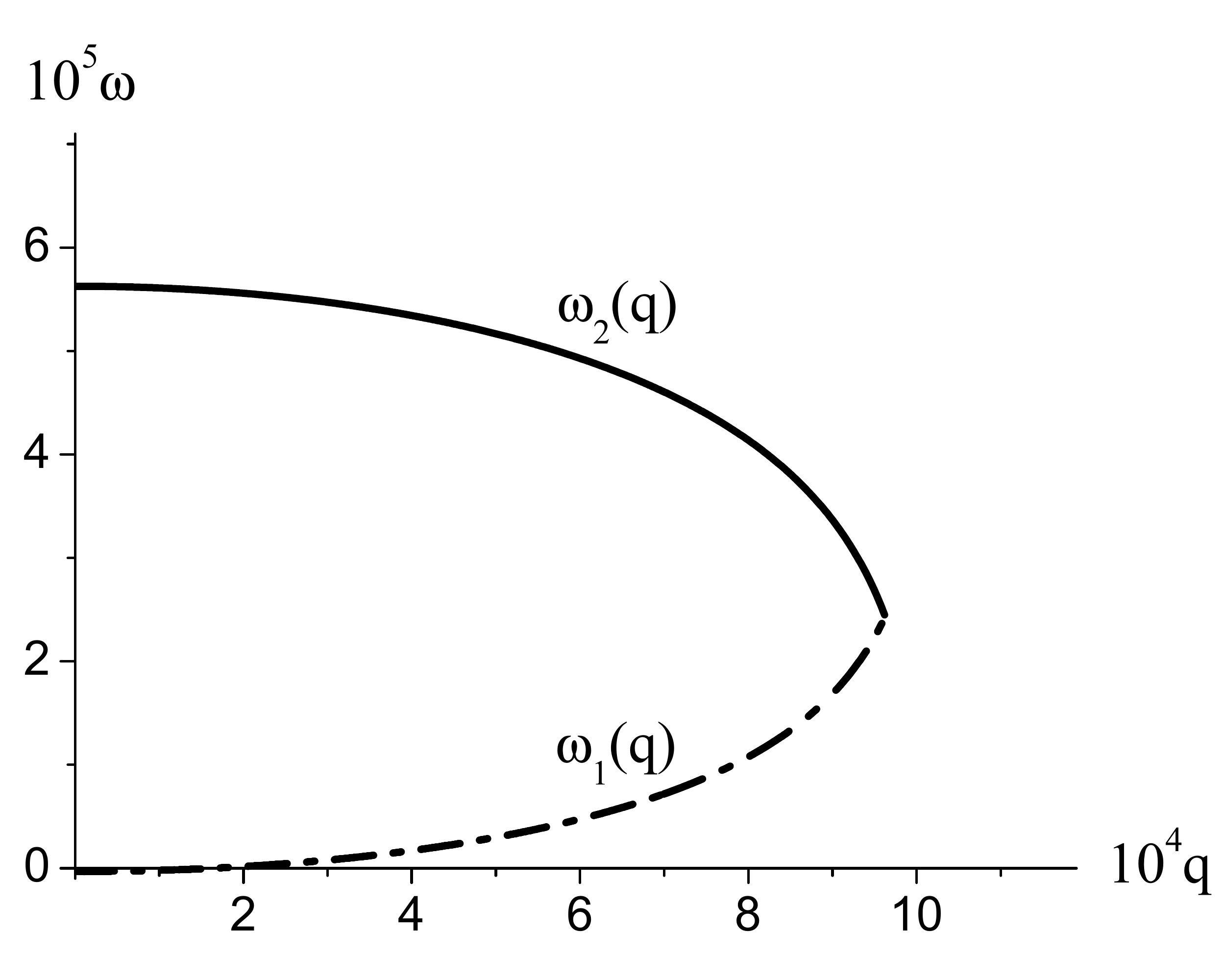}
 \caption{\label{MS+Cahn}{Dispersion curves of both unstable modes $\omega_1(q)$ and $\omega_2(q)$, arising for the parameter values $p=100$, $\gamma=0.01$, $v=0.05$.}}
 \end{figure}
 \end{center}

With that, a numerical evaluation of Eq. (\ref{dispersion}) in the unstable regime of both modes leads to the
dispersion curves, shown in Figs. 8, 9, and 10 for real-valued rates $\omega_1$, and $\omega_2$. They consecutively refer to the Mullins-Sekerka-like instability, the Cahn instability, and to the superposition of both instabilities. The complete loop in Fig. 10 shrinks to the point $q=\omega=0$, when the stability limit $F'(v)=0$ is approached, and it opens again in the stable region $\omega<0$, when $F'(v)$ grows from zero to positive values. Qualitatively, the same behavior has been found by Braun et al. \cite{Braun} in a phase-field model of solidification.

A convenience of the capillary-wave description (\ref{I-Hamiltonian}) and (\ref{I-dynamics}) is that it intuitively invites for applications to toy models with piece-wise linear functions $U(\zeta)$ and $D(\zeta)$. A similar strategy has been used in the context of grain-boundary motion by Cahn \cite{Cahn}, and later, more extensively, by Hillert \cite{Hillert}. In the present case of binary-alloy solidification all consecutively considered families of such models have the equilibrium-concentration profile

\begin{equation}\label{tilt}
U(\zeta)=\Theta(\zeta+\delta)\Theta(\delta-\zeta)\frac{\zeta+\delta}{\delta}+2\Theta(\zeta-\delta)\,\,.
\end{equation}
As shown in Fig. 11, this expression linearly interpolates between the two bulk phases, and for $\delta=1$ mimics the the preceding model $U(\zeta)=\Phi_E(\zeta)+1$ whereas for $\delta\rightarrow 0$ it approaches the sharp-interface model $U(\zeta)=2\Theta(\zeta)$. If Eq. (\ref{tilt}) is attended by a uniform diffusion coefficient $D(\zeta)=1$, Eqs. (\ref{I-G-C}) lead to the expression

\begin{equation}\label{G-tilt}
G_F(v)=\gamma\bigl[1-2v\delta-\exp{(-2v\delta)}\bigr]\frac{1}{(v\delta)^2}\,\,,
\end{equation}
and, as a consequence, to the the behavior

\begin{equation}\label{tilt-anomaly}
F'(0)=\frac{1}{p}-\frac{1}{p_c}\,\,\,,\,\,\,p_c\equiv\frac{3}{2\,\delta\gamma}\,\,.
\end{equation}
From this result we see that the Cahn anomaly $F'(0)<0$ exists for finite values of $\delta$, but disappears in the sharp-interface approximation. This is obviously the reason why the anomaly is even not mentioned in discussions, based on sharp-interface models, supported by local-equilibrium boundary conditions for the concentration field. As mentioned below Eq. (\ref{anomal}), the anomaly is not excluded in the Aziz-Boettinger approach \cite{AB} where an internal structure of the interface region was effectively taken into account.

\begin{center}
\begin{figure}
 \includegraphics[width=8cm]{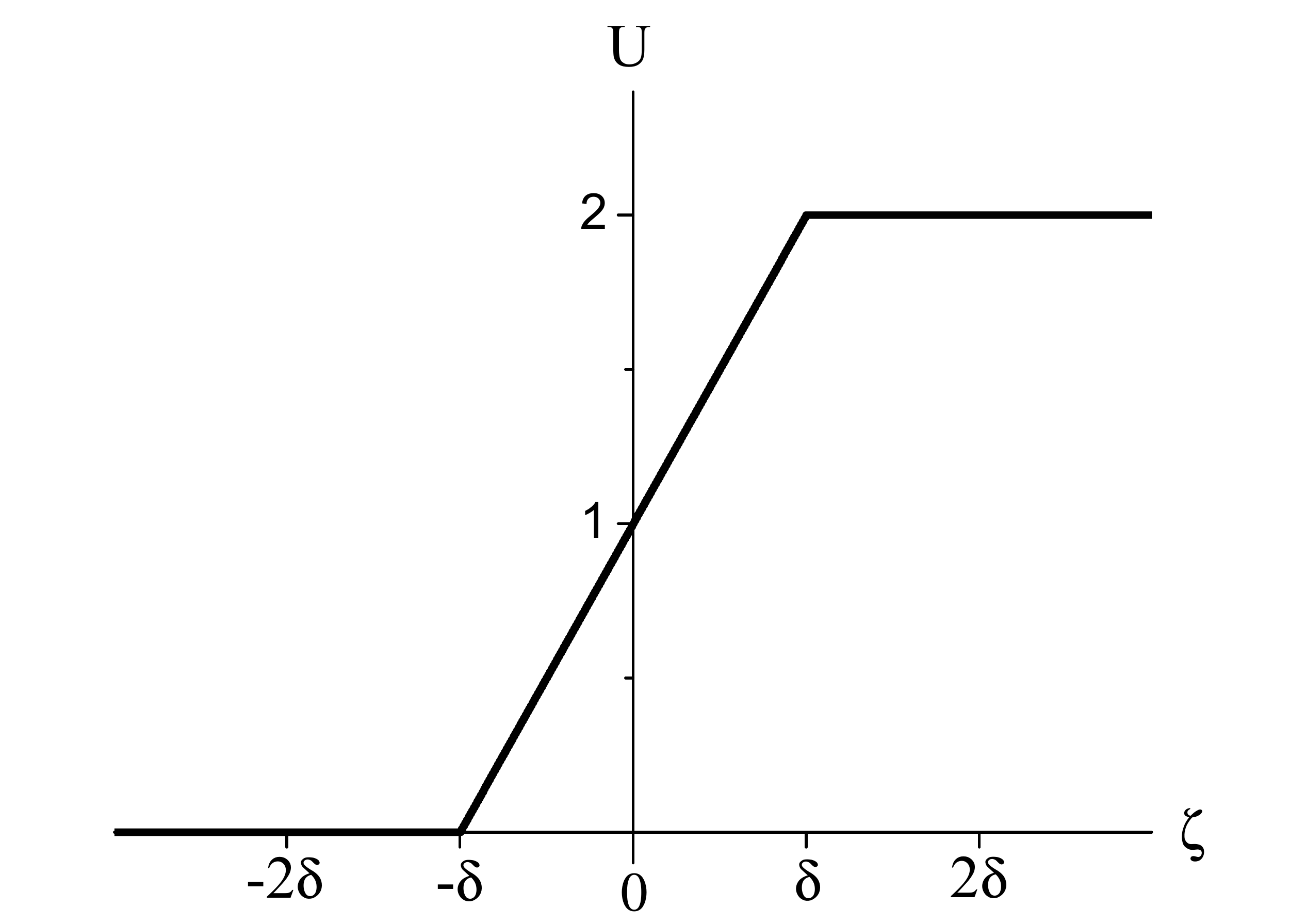}
 \caption{\label{tilt-model}{The potential $U(\zeta)$ which, with a free parameter $\delta$, linearly interpolates between the liquid and solid phases. The limit $\delta\rightarrow 0$ defines the sharp-interface limit in our approach.}}
 \end{figure}
 \end{center}

In Ref. \cite{PRB} we have considered a set of model systems which included the profile (\ref{tilt}), but was complemented by a diffusion coefficient

\begin{equation}\label{step-diffusion}
D(\zeta)=\Theta(\zeta+\delta)\,\,.
\end{equation}
Whereas for finite $\delta$ the Cahn anomaly again appeared, we recovered in the limit $\delta\rightarrow 0$ the instability, discussed by Misbah et al. \cite{MMT} which is a kind of extension of the Mullins-Sekerka instability \cite{MS} from the diffusion- into the kinetics-limited regime.

As a final application we now consider a class of models where the equilibrium concentration (\ref{tilt}) is attended by the diffusion coefficient

\begin{equation}\label{tilt-diffusion}
D(\zeta)=\frac{1}{2}\,U(\zeta)
\end{equation}
which, consequently, also linearly interpolates between the two bulk phases. This behavior mimics the result of numerical calculations, derived in Refs. \cite{Laird}, and \cite{Binder} from the mean-square particle displacements. Adopting the form (\ref{tilt-diffusion}), one easily verifies that the steady-state equation

\begin{equation}\label{tilt-C-U}
C_F'(\zeta)+\frac{v}{D(\zeta)}C_F(\zeta)=U'(\zeta)
\end{equation}
has the, again linearly interpolating, solution 

\begin{eqnarray}\label{tilt-C-F}
C_F(\zeta)&=&\Theta(\zeta+\delta)\Theta(\delta-\zeta)\frac{1}{1+2v\delta}\frac{\zeta+\delta}{\delta}\\&+&
\Theta(\zeta-\delta)\frac{2}{1+2v\delta}\exp{[-v(\zeta-\delta)]}\,\,.\nonumber
\end{eqnarray}
In view of the special value $C_F(0)=1/(1+2v\delta)$ we observe that the partition coefficient $K(v)$ again has the form (\ref{Aziz}), however, with the new reference velocity

\begin{equation}\label{delta-V*}
V^*=\frac{1}{2\delta}\frac{V_D}{K_E}\,\,.
\end{equation} 
In the measurements \cite{SA} of $K(v)$ the quantity $\delta$ may be used as a fitting parameter. Moreover, from the first formula in Eqs. (\ref{I-G-C}) and the result (\ref{tilt-C-F}) we obtain

\begin{equation}\label{tilt-drag}
G_F(v)=-\,\gamma\,\frac{2}{1+2v\delta}
\end{equation}
which, via the force-balance relation in Eqs. (\ref{balance}), implies

\begin{equation}\label{tilt-F(v)}
F(v)=\frac{v}{p}-\frac{v}{p_c}\,\frac{1}{1+2v\delta}\,\,\,,\,\,\,p_c\equiv\frac{1}{4\delta\gamma}\,\,.
\end{equation}
This function is convex in the sense $F''(v)\ge 0$, and has a minimum at

\begin{eqnarray}\label{tilt-minimum}
v_m&=&\frac{1}{2\delta}\left(\sqrt{\frac{p}{p_c}}-1\right)\,\,,\\
F_m&=&-2\gamma\left(1-\sqrt{\frac{p_c}{p}}\,\right)^2\,,\nonumber
\end{eqnarray}
so that the Cahn anomaly $F'(v)<0$ occurs in the whole regime $v<v_m$. By means of Eq. (\ref{Undercooling}) the value $F_m$ corresponds to the temperature 

\begin{equation}\label{T-maximum}
T_m=T_S+\frac{1}{2}\left(1-\sqrt{\frac{p_c}{p}}\,\right)^2(T_L-T_S)
\end{equation}
where all temperatures refer to the density $C_S$. The value $T_m$ limits the temperature range, up to which a meta-stable kinetics-limited solidification is possible inside the diffusion-limited regime. In the constant-miscibility-gap approximation the line $T_m(C)$ is, due to Eq. (\ref{T_0-line}), located below the Baker-Cahn line $T_0(C)$, and approaches it in the limit $p\rightarrow\infty$.

If Eq. (\ref{tilt-F(v)}) is rewritten in terms of the scaled variables

\begin{equation}\label{f-w}
w\equiv\frac{1}{2\gamma}\frac{V}{V_c}\,\,\,,\,\,\,f\equiv\frac{1}{2\gamma}F\,\,,
\end{equation}
it assumes the simple form

\begin{equation}\label{f(w)}
f(w)=w-\frac{p}{p_c}\,w\left[1+\frac{p}{p_c}\,w\right]^{-1}
\end{equation}
where only the ratio $p/p_c$ enters as a tunable parameter. Elimination of this parameter from Eqs. (\ref{tilt-minimum}) leads to the trajectory

\begin{center}
\begin{figure}
 \includegraphics[width=8cm]{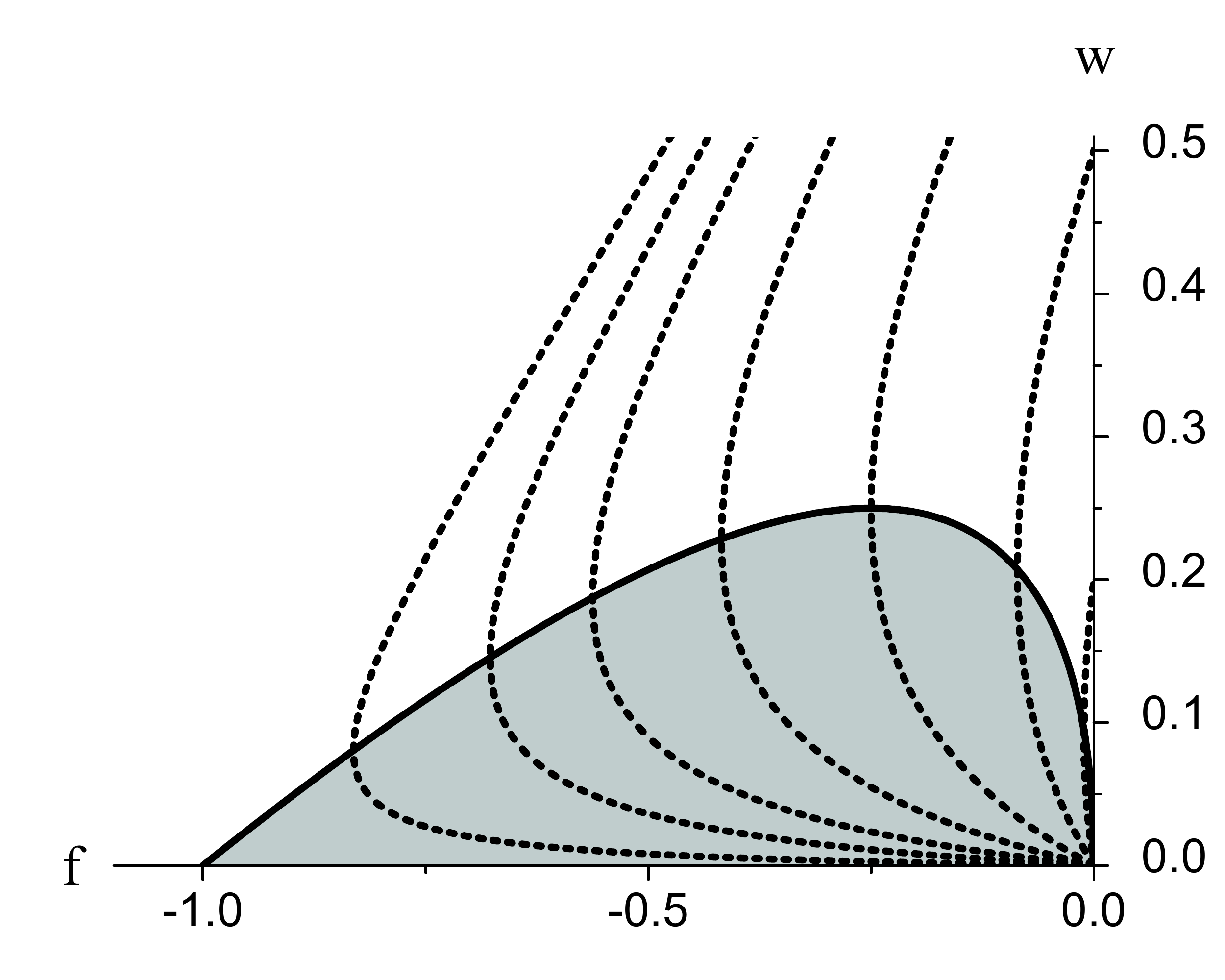}
 \caption{\label{kin-spin}{Kinetic spinodal line $w(f)$, enclosing the region where the Cahn instability $f'(w)<0$ occurs in the $f,w$-plane. The limiting point $w(-1)=0$ corresponds to the adiabatic limit in the one-component system. Also shown are some curves $f(w)$ for different values of the parameter $p/p_c$.}}
 \end{figure}
 \end{center}

\begin{equation}\label{trajectory}
w=\sqrt{-f}\,\left(1-\sqrt{-f}\right)
\end{equation}
in Fig. 12, enclosing a regime of unstable behavior in the sense $f'(w)<0$. The outside region, up to the line $f=0$, is a a regime of meta-stable solidification. Accordingly, the trajectory may be considered as a spinodal line of kinetic origin which starts at the origin $w=f=0$ for $p=p_c$, and approaches the point $w=0,f=-1$ for $p\rightarrow\infty$, corresponding to the limit $D_L\rightarrow 0$.  

Remarkably, the result (\ref{trajectory}) precisely agrees with an analogous stability limit, derived by Umantsev for the non-isothermal solidification of a pure one-component substance \cite{Umantsev}. Since, in this case, the diffusion constant $D_L$ has to be replaced by the heat-diffusion coefficient $D_T$, the above-mentioned limit $D_T\rightarrow 0$ corresponds to the process of adiabatic solidification.

\section{Discussion}

The most remarkable implication of our approach is the existence of the Cahn instability $\omega_2(q)>0$ which arises in case of the anomaly $F'(v)<0$. A regime with this behavior occurs in all considered
models, except in the sharp-interface limit where only the Mullins-Sekerka instability survives. The behavior $F'(v)<0$ even exists at the origin $v=0$ which has also been noticed in many approaches, based on a phase-field description. 

In view of the anomalous behavior $F'(0)<0$ one may question the consistency of our model, and also that of standard phase-field models, with the basic principle of a positive entropy production. It has been shown, however, by Bi and Sekerka \cite{BS} that, for a general class of phase-field models, the entropy production is positive. Applied to our model, and using our notations, their expression for the entropy production can be written in the form

\begin{equation}\label{S-production}
\Pi=\int d^3r\left[\frac{1}{\Gamma}\,(\partial_t\Phi)^2+ \frac{\mathcal{D}}{\kappa}\left(\nabla\frac{\delta\mathcal{H}}{\delta\mathcal{C}}\right)^2\right]
\end{equation}
which obviously is positive. In order to ensure that the expansion scheme, underlying our capillary-wave model, does not endanger this property, we apply Eq. (\ref{S-production}) to the one-dimensional steady-state solidification, replacing everywhere $\Phi_F(\zeta)$ by the static profile $\Phi_E(\zeta)$. The result for the entropy production per unit area then reads in dimensionless units

\begin{eqnarray}\label{S-prod}
&&\pi=\int d\zeta\left[\frac{v^2}{p}\,(\partial_\zeta\Phi_E)^2+D\gamma(C_F'-U')^2\right]
\nonumber\\&&\,\,\,\,\,=v\left[\frac{v}{p}-G_F(v)\right]=v[F(v)-G_F(0)]
\end{eqnarray}
where we have used the expression (\ref{mov-kink}) for $\Phi_E(\zeta)$, the representation of $G_F(v)$ in the first line of Eqs. (\ref{I-G-C}), the relation (\ref{C_F-U}), and the force balance in Eqs. (\ref{balance}). In view of Eq. (\ref{G-properties}) the quantity (\ref{S-prod}) again is positive which remains true in the expanded form

\begin{equation}\label{P-anomaly}
\pi\approx v[2\gamma+vF'(0)]
\end{equation}
and, consequently, demonstrates that the strange-looking behavior $F'(0)<0$ is not in conflict with the second law of thermodynamics.

To complete our presentation, we add a few estimates which border the range of applicability of our approach. First of all, it is clear that the two small parameters $\gamma$ and $v/p$ of our approach can be tuned by the miscibility gap $\Delta C$ and by the driving force $F$, respectively. From Eqs. (\ref{kappa(T)}) and (\ref{dimensionless}) we know that

\begin{equation}\label{gamma-Delta C}
\gamma\equiv-\,\frac{\xi L}{4\sigma}\,\left(\frac{\partial C_L}{\partial T}\right)^{-1}\frac{\Delta C}{T_M}\,\,.
\end{equation}
According to Turnbull \cite{Turnbull} there is a correlation between the surface tension and the latent heat of the form
\begin{equation}\label{turnbull}
\sigma=C_T\,L\,a 
\end{equation}
where $a$ measures the average atomic distance, and, for essentially all metals, $C_T\approx 0.45$. Assuming $\xi\approx 1.8\,a$, and expressing the miscibility gap by the temperature gap

\begin{equation}\label{temp-gap}
\Delta T\equiv T_L(C_S)-T_S(C_S)\,\,, 
\end{equation} 
Eq. (\ref{gamma-Delta C}) reduces to 

\begin{equation}\label{gammaprox}
\gamma\approx\frac{\Delta T}{T_M}\,\,. 
\end{equation} 
For the values $\Delta T=10\,K$ and $T_M=1000\,K$ this yields $\gamma\approx 0.01$ which is sufficiently small, and identical to the value, underlying Figs. 9 and 10. The value $p=100$, also used in these figures, has been adopted from Ref. \cite{AK}, and, in the rapid growth regime $v\approx 1$ implies $v/p\approx 0.01$ for our second expansion parameter.

As a final remark we point out that the capillary-wave model, defined by Eqs. (\ref{I-Hamiltonian}) and (\ref{I-dynamics}), is amenable to a variety of generalizations without going back to a phase-field          description. One generalization in the surface part of the effective Hamiltonian (\ref{I-Hamiltonian}) is the replacement of the integral expression by the exact area of the interface which in the equation of motion leads to the appearance of the full mean curvature. Simultaneously, one may also incorporate an anisotropy of the surface tension. Further generalizations are the inclusion of the energy density or other bulk fields which then generates a number of cross couplings in the Hamiltonian. Parallel to this, Eqs. (\ref{I-dynamics}) will be replaced by an enlarged set of equations of motion which in general is equipped with a matrix of Onsager coefficients. A consistent discussion of the directional solidification of a dilute binary alloy requires a coupling of the energy density to the interface position which we will consider in a forthcoming paper.

\acknowledgments

We are grateful to  Efim Brener for calling our attention to the consistency problem, examined in our discussion. A. L. K. wants to express his gratitude to the University of D\"usseldorf for its warm hospitality. This work has been supported by the DFG under BA 944/3-3, and by the RFBR under N10-02-91332.

\end{document}